\documentclass[a4paper,prd,superscriptaddress,11pt,noshowkeys,notitlepage,nofootinbib]{revtex4}

\usepackage[utf8]{inputenc}
\usepackage[T1]{fontenc} % if needed
\usepackage[list=true]{subcaption}
\usepackage{slashed}
\usepackage{multirow}
\usepackage{enumitem}
\usepackage{amsmath, amssymb}
\usepackage{graphicx}
\usepackage{dsfont}
\usepackage{caption}
\usepackage[normalem]{ulem}

\newcommand{\beq}{\begin{equation}}
\newcommand{\eeq}{\end{equation}}
\newcommand{\ba}{\begin{array}}
\newcommand{\ea}{\end{array}}
\newcommand{\bea}{\begin{eqnarray}}
\newcommand{\eea}{\end{eqnarray} }
\newcommand{\bal}{\begin{align}}
\newcommand{\eal}{\end{align}}

\begin{document}

\title{\boldmath Two-loop renormalisation of gauge theories in $4D$ Implicit Regularisation and connections to dimensional methods}

%% %simple case: 2 authors, same institution
%% \author{A. Uthor}
%% \author{and A. Nother Author}
%% \affiliation{Institution,\\Address, Country}

% more complex case: 4 authors, 3 institutions, 2 footnotes
\author{A. Cherchiglia}\email[]{adriano.cherchiglia@ufabc.edu.br}
\author{D. C. Arias-Perdomo}\email[]{carolina.perdomo@ufabc.edu.br}
\affiliation{CCNH, Universidade Federal do ABC, \\ 09210-580 , Santo Andr\'e - SP, Brazil}
\author{A. R. Vieira}\email[]{alexandre.vieira@uftm.edu.br}
\affiliation{Universidade Federal do Tri\^angulo Mineiro, \\Iturama, MG 38280-000, Brazil}
\author{M. Sampaio}\email[]{marcos.sampaio@ufabc.edu.br}
\affiliation{CCNH, Universidade Federal do ABC, \\ 09210-580 , Santo Andr\'e - SP, Brazil}
\author{and B. Hiller}\email[]{brigitte@fis.uc.pt}
\affiliation{CFisUC, Department of Physics, University of Coimbra,\\ P-3004-516 Coimbra, Portugal.}

% The "\note" macro will give a warning: "Ignoring empty anchor..."
% you can safely ignore it.

%\affiliation{(a) CCNH, Universidade Federal do ABC, \\ 09210-580 , Santo Andr\'e - SP, Brazil}
%\affiliation{(b) Universidade Federal do Tri\^angulo Mineiro, \\Iturama, MG 38280-000, Brazil}
%\affiliation{(c) CFisUC, Department of Physics, University of Coimbra,\\ P-3004-516 Coimbra, Portugal.}

%% e-mail addresses: one for each author, in the same order as the authors
%\emailAdd{adriano.cherchiglia@ufabc.edu.br}
%\emailAdd{carolina.perdomo@ufabc.edu.br}
%\emailAdd{alexandre.vieira@uftm.edu.br}
%\emailAdd{marcos.sampaio@ufabc.edu.br}
%\emailAdd{brigitte@fis.uc.pt}

\begin{abstract}

We compute the two-loop $\beta$-function of scalar and spinorial quantum electrodynamics as well as pure Yang-Mills and quantum chromodynamics using the background field method in a fully quadridimensional setup using Implicit Regularization (IREG). Moreover, a thorough comparison with dimensional approaches such as conventional dimensional regularization (CDR) and dimensional reduction (DRED) is presented. Subtleties related to Lorentz algebra contractions/symmetric integrations inside divergent integrals as well as  renormalisation schemes are carefully discussed within IREG where the renormalisation constants are fully defined as basic divergent integrals to arbitrary loop order. Moreover, we confirm the hypothesis that momentum routing invariance in the loops of Feynman diagrams implemented via setting well-defined surface terms to zero deliver non-abelian gauge invariant  amplitudes within IREG just as it has been proven for abelian theories.
%\AC{talvez suavizar? Uma crítica do referee era que não mostramos nesse paper que IREG respeita QCD em geral. Sugestão:} Moreover, we confirm the hypothesis that momentum routing invariance in the loops of Feynman diagrams implemented via setting well-defined surface terms to zero deliver non-abelian gauge invariant  amplitudes within IREG just as it has been proven for abelian theories.}}
\end{abstract}

\maketitle
%\flushbottom

\section{Motivations}
\label{sec:intro}

Unravelling physics beyond the standard model (SM) has entreated theoretical predictions for particle physics precision observables beyond next-to-leading-order ($NLO$). Such predictions rely on involved Feynman diagram calculations to evaluate scattering amplitudes both in the SM  and its extensions. Theoretical models beyond the SM (BSM) can be constructed, for instance, as an extension in the Higgs sector by either changing the number of scalar multiplets or considering the Higgs boson as a composite particle -- the so called Composite Higgs Models \cite{HSEC,CHM}. Supersymmetric and dark matter extensions have also been considered in order to explain SM deviations from experimental results \cite{REVIEW} in electroweak precision observables (EWPO) which are known with an accuracy at the per cent level or better \cite{EWPO,ADRISTO1,EBOLI}. On the other hand, precise measurements and calculations of known particles and interactions are just as important to validate, redress, or refute new models. Also, in order to evade from unphysical scale dependence at low order, higher order terms are needed to smooth out such dependence  in the resulting, more accurate, predictions. For example, a full $N^3LO$ calculation for  QCD corrections to gluon-fusion Higgs boson production was performed in  \cite{ANAST1} at center-of-mass energy 13 $TeV$. The considerably low residual theoretical uncertainty ($\approx 5-6 \%$) and small sensitivity to scale variation ($\approx 2 \%$ ) superseded earlier results below $N^3LO$.  Because experimental uncertainties are expected to drop below the accuracy of theoretical data, as expected from future experimental measurements at the Future Circular Collider  (FCC-$e^-e^+$) \cite{HTTPS}, $QCD$ theoretical  uncertainties ought to be reduced at many levels so physics BSM can be ultimately ascertained.

Ultraviolet (UV) and infrared (IR) divergences are ubiquitous beyond leading order in S-matrix calculations and must be judiciously removed in order to automated computation codes for the evaluation of Feynman amplitudes. 
As a by-product of such subtractions, there remain residual dependencies on renormalisation ($\lambda_R$) and factorisation ($\lambda_F$) scales in the perturbative series that describes a physical observable. The dependence on such scales is expected  diminish  after higher terms are taken into account and, at a given order, may in principle be minimised to yield a result the least sensitive  to variations in the unphysical parameters \cite{STEVEN}. However, the problem of scale setting has been studied extensively and there is no consensus on a procedure valid in general. For a recent account see \cite{Farrukh}.

For typical collider observables, due to the high multiplicity of jets, the real radiation is subjected to intricate
phase-space constraints and calculations are often performed numerically. In such cases the cancellation of soft and collinear IR divergences becomes more involved and precise evaluations of parton distribution functions (PDF's), needed to calculate cross-sections with hadron beams, become mandatory \cite{PDF}.  A general cross section in $QCD$ usually includes short and long-distance behaviour and thus it is not computable directly in perturbation theory. This remarkable problem was first addressed by Weinberg in his pioneering work on $QED$ and quantum gravity \cite{WEINBERG}. At a given order $\alpha_s^n$ and momentum transfer $Q$ there appear large logarithms such as renormalisation and factorisation logs $\alpha_s^2 \ln^n (Q^2/\lambda_R^2)$, $\alpha_s^2 \ln^n (Q^2/\lambda_F^2)$, as well as high energy logs and Sudakov logs \cite{LOG}. Renormalisation group (RG) logs are resummed via RG evolution equations whereas Sudakov logs originate from IR and collinear singularities and may be resummed through exponentiation of IR and collinear poles. Such resummations at and beyond next-to-leading-log ($NLL$)  assure the validity of the perturbative series, leading to non-perturbative contributions to high energy cross sections. For those resummations, it is crucial to rely on
factorisation theorems \cite{FACTOR1,GARDI,FACTORSTOCK}. Factorisation properties  separate the dynamics at different energy scales. For instance, for $n$ external partons of momenta $p_i$ in the high energy limit, an amplitude factorizes as \cite{FACTORSTOCK}
\bea
{\cal{M}}_n \Bigg(\frac{\mu}{\lambda_{IR}}, \frac{p_i}{\lambda_{UV}}, \alpha_s (\lambda_{UV})\Bigg) = Z \Bigg(\frac{\mu}{\lambda_{IR}}, \frac{p_i}{\lambda_{IR}},\alpha_s (\lambda_{IR})\Bigg)H_n\Bigg(\frac{p_i}{\lambda_{UV}},
\frac{\lambda_{IR}}{\lambda_{UV}}, \alpha_s (\lambda_{UV})\Bigg),
\eea
where $H_n$ is the UV renormalized piece and $Z$ contains the IR soft and collinear divergences expressed here by $\mu \rightarrow 0$. The factor $Z$ itself obeys an RG-like equation which gives rise to anomalous dimensions. The latter  can be used to resum Sudakov logs, non-abelian exponentiation of differential cross-sections \cite{FRENKEL}, as well as to understand the mapping between IR  onto UV divergences of operators in effective field theory \cite{FACTOR1}. 

In order to tackle the problems discussed above, the choice of a regularisation scheme for UV or IR divergences in Feynman amplitudes matters from both conceptual and practical aspects. In theory the choice of regularisation is unphysical but in practice it is paramount to both studying anomalies in perturbative quantum field theory and automatising loop calculations.
In recent years, novel schemes have been proposed aiming at improving or even obliterating conventional dimensional regularisation (CDR)\cite{REGDIM}. In dimensional specific models, such as chiral \cite{GAMMA5}, topological \cite{CHERNSIMONS} or supersymmetric quantum field theories \cite{REGSUSY}, CDR  needs caveats or even explicit modifications  such as the dimensional reduction scheme (DRED) \cite{SIEGEL,SIEGEL2}. On the other hand, such modifications on CDR can lead to  
additional terms at the Lagrangian level, such as the $\epsilon$-scalar particles \cite{EPSILONSCALAR,EPSILONSCALAR-BETA}, in order to satisfy Ward identities and account for amplitude factorisation and other renormalisation group properties of the model. The downsides are that besides the Feynman rules become more involved, such fictitious particles are not protected by gauge invariance and require  coupling $\alpha_\epsilon, \alpha_{4 \epsilon} \ne \alpha_s$, and thus their own $\beta$ functions, in order to preserve unitarity. 
% \AC{[REMOVE?] Another dimensional approaches, with focus on numerical evaluations, aim to generalize the OPP algorithm \cite{OPP} to NNLO (and beyond). This relies on the identification of the some called rational terms, whose master formula at two-loop level became recently available \cite{ZOLLER}.}

Alternatively, some schemes that operate only on the physical dimension of the underlying model aside from energy cut-offs have been constructed. Amidst them, implicit regularisation (IREG) was constructed for the purpose of shedding light on theoretical aspects of regularisation dependent quantum corrections as well as providing novel
analytical and hopefully numerical calculational methods \cite{ZURICH}. Transition rules to other schemes is particularly perspicuous within IREG.

In this contribution, we focus on the UV renormalisation of scalar/spinorial QED, pure Yang-Mills and QCD to two loop order within IREG. Working entirely in four dimensions, we calculate the $\beta$-functions for the gauge coupling constant to two-loop order %using the background field method (BFM) 
in a minimal subtraction scheme of the basic divergent integrals (BDI's) in internal momenta. BDI's contain a natural renormalisation scale $\lambda_R \equiv \lambda$ and are absorbed into renormalisation constants whose counterparts in DRED and CDR are provided. By definition, the subtraction scheme adopted by us is
%Such scheme is
%, by definition, 
mass independent, implying that the first two 
coefficients of the $\beta$-function of gauge couplings are universal (renormalisation scheme independent) \cite{TARRACH}. Therefore, we reproduce the well-known results obtained in the context of dimensional regularization within the minimal subtraction scheme.

Moreover, our results show compliance with gauge symmetry lending support to the conjecture that momentum routing invariance (MRI) in Feynman diagrams (enforced by setting well-defined surface terms to zero)  delivers gauge invariant amplitudes. Such a conjecture was proved to be valid in the abelian gauge theories to all orders in perturbation theory \cite{PRD2012,ALEXANDRE}. No modifications at Lagrangian level are necessary as IREG operates in the physical dimension. For instance, no analog of $\epsilon$-scalar field needed in schemes such as DRED and four-dimensional helicity (FDH) \cite{FACTORSTOCK,FDH} to assure unitarity and proper cancellation of divergences is introduced in IREG as a matter of principle as it operates in the physical dimension. We show nevertheless, that in this particular calculation, $\epsilon$-field contributions cancel out in DRED. It is noteworthy that $\epsilon$-scalars are crucial to understand IR structure in DRED or FDH schemes.

In the next section we provide a brief panorama over regularisation schemes with focus upon IREG rules that will be used throughout this contribution. We also discuss some subtleties related to Lorentz algebra contractions/symmetric integrations inside divergent integrals, and present a dedicated analysis of the correspondence among IREG and dimensional methods. In section \ref{sec:results} we study a series examples at two-loop order, aiming to compute the first two coefficients of the gauge coupling $\beta$ function in scalar/spinorial QED, pure Yang-Mills, and QCD. A careful comparison with CDR and DRED is performed. Finally, we conclude in section \ref{sec:conclusions}, and present a series of appendixes containing some technical details of the calculations. 

\section{Survey of regularisation schemes and IREG rules}
%\label{sec:survey}

In CDR \cite{REGDIM} the vector bosons are treated in $d= 4 - 2\epsilon$ dimensions. This formulation cannot be consistently applied to dimensional specific models such as chiral, topological and supersymmetric quantum field theories as discussed in the introduction. Some variants of CDR such as the 't Hooft-Veltman method (HV) \cite{MAISON}, dimensional reduction (DRED)\cite{SIEGEL,SIEGEL2,FACTORSTOCK,REGSUSY,EPSILONSCALAR} and Four Dimensional Helicity (FDH) \cite{FDH} have been developed. Moreover some scattering amplitudes in $QCD$ are regularisation dependent \cite{REGDEPQCD} mainly due to the interplay between IR and UV divergences to yield finite results. In dimensional methods both UV and IR infinities appear as poles  $1/\epsilon^n$. An unclear distinction between the origin of such poles can lead to ambiguities such as the nature of radiative contributions to supersymmetric Yang-Mills $\beta$-functions  \cite{MAS}. We may schematise dimensional methods as in table  \ref{tab:DIM}. 
\begin{table}[h!]
	\begin{center}
		\begin{tabular}{l|cccc}
		&CDR&HV&FDH&DRED\\
			\hline
			internal gluon  &$\hat{g}^{\mu\nu}$&$\hat{g}^{\mu\nu}$&
			$g^{\mu\nu}$&$g^{\mu\nu}$\\
			external gluon&$\hat{g}^{\mu\nu}$&$\bar{g}^{\mu\nu}$&
			$\bar{g}^{\mu\nu}$&$g^{\mu\nu}$
		\end{tabular}
	\end{center}
\caption[Let $n$ be the physical dimension. In CDR and HV, gluons are regularised in  $d$ dimensions with metric tensor $\hat{g}_{\mu \nu}$ (QdS). The quasi-n-dimensional space has metric  $g_{\mu \nu}$ (QnS) and the original $n$-dimensional space is denoted by  $\bar{g}_{\mu \nu}$ (nS). A complementary   $\varepsilon$-dimensional tensor $\tilde{g}_{\mu \nu}$  must be introduced such that $g_{\mu \nu}=\hat{g}_{\mu \nu} + \tilde{g}_{\mu \nu}$. Thus $QnS=QdS \oplus Q \varepsilon S$. Mathematical consistency and gauge invariance require  $QnS \supset QdS \supset nS$ and prohibit to identify $g_{\mu \nu}$ with $\bar{g}_{\mu \nu}$ \cite{REGSUSY}.]{\label{tab:DIM} Let $n$ be the physical dimension. In CDR and HV, gluons are regularised in  $d$ dimensions with metric tensor $\hat{g}_{\mu \nu}$ (QdS). The quasi-n-dimensional space has metric  $g_{\mu \nu}$ (QnS) and the original $n$-dimensional space is denoted by  $\bar{g}_{\mu \nu}$ (nS). A complementary   $\varepsilon$-dimensional tensor $\tilde{g}_{\mu \nu}$  must be introduced such that $g_{\mu \nu}=\hat{g}_{\mu \nu} + \tilde{g}_{\mu \nu}$. Thus $QnS=QdS \oplus Q \varepsilon S$ . Mathematical consistency and gauge invariance require  $QnS \supset QdS \supset nS$ and prohibit to identify $g_{\mu \nu}$ with $\bar{g}_{\mu \nu}$ \cite{REGSUSY}.}
\end{table}

Grosso modo, a quasi-4-dimensional gauge field may be decomposed as 
$
A^\mu = \hat{A}^\mu + \tilde{A}^\mu,
$
where $\hat{A}^\mu$ is a  $d$-dimensional gauge field and  $\tilde{A}^\mu$ is a {\it{scalar}} of multiplicity $N_\epsilon = 2 \epsilon$ \cite{BROGGIO}. This amounts to adding up to the original Lagrangian a term ${\cal{L}}_\epsilon$ that contains evanescent Yukawa couplings between $\epsilon$-scalars and quarks of strength $\lambda_\epsilon$ and quartic $\epsilon$-scalar vertices with strength $\lambda_{4 \epsilon}$ \cite{EPSILONSCALAR}.
The resulting schemes such as DRED and FDH are crucial to study models with supersymmetry. The latter is explicitly broken in  CDR as gauge bosons are considered as quantities with $d$ components whereas gauginos remain $4$-dimensional which means an unbalance between fermionic and bosonic degrees of freedom. Symmetry restoring conterterms may be added order by order in perturbation theory in the renormalised theory \cite{KRAUSSTOCK,KRAUS}. However, besides being a striking additional complication, we cannot discard possible anomalous symmetry breakings in supersymmetric gauge theories in general \cite{REGSUSY}. In this sense, DRED has been shown to be an invariant supersymmetric scheme to minimal supersymmetric gauge models up to two loop order using the quantum action principle \cite{SUSYQCD}. 

For an efficient computational code to evaluate Feynman amplitudes beyond leading order\footnote{For a recent application in the framework of dimensional approaches see, for instance, \cite{ZOLLER}}, UV and IR divergences ought to be subtracted by a scheme that respects unitarity, causality, and symmetries. %\footnote{From the point of view of numerical evaluation in dimensional schemes, efficient approaches aim to generalize the OPP algorithm \cite{OPP} to NNLO (and beyond). This relies on the identification of the some called rational terms, whose master formula at two-loop level became recently available \cite{ZOLLER}.
%}
The subtraction of nested and overlapped divergences should respect the Bogolyubov recursion relations, locality, and the BPHZ algorithm \cite{BPHZ}. The techniques can be divided into two categories \cite{ZURICH}:
\begin{enumerate}
	\item ``to d'': CDR (1972), HV (1977); DRED (1979), FDH (1992), Six-dimensional formalism SDF (2009) \cite{SDF}, Four dimensional Formalism
	FDF (2014) \cite{FDF}. 
	\item ``or not to d'': Higher covariant derivative regularization (1971) HD \cite{HDR}; Differential regularisation/renormalisation DIFR (1992) \cite{FREEDMAN}, Implicit regularisation IREG (1998) \cite{PRIM2}, Loop regularisation LORE (2003) \cite{LORE}, Four-Dimensional regularisation/renormalisation  FDR (2012) \cite{FDR}, Four-Dimensional Unsubtraction FDU (2016) \cite{FDU}.
\end{enumerate}

In this contribution we will be mainly focused on the latter set, in particular the IREG framework whose rules and state of art we present in the next subsection.

\subsection{State of the art and rules of IREG}
\label{sec:rules}

The extraction of  UV divergent basic divergent integrals in IREG can be effected  in consonance with  Bogoliubov's recursion formula \cite{ADRIANO} complying with
unitarity, Lorentz invariance and locality; a proof-of-concept calculation has shown how renormalisation functions of  scalar theories are obtained beyond one loop order. In \cite{PRD2012,ALEXANDRE} it was shown that IREG respects abelian gauge symmetry to $N$-loop order in perturbation theory, in a constrained version in which  surface terms (ST's) are set to vanish. Such regularisation dependent  ST's serve as a tag for momentum routing dependence in the loops of Feynman diagrams and therefore an important connection between momentum routing invariance and gauge symmetry was established. ST's also parametrise  finite and arbitrary parameters in Feynman diagram calculations which  may  be fixed on symmetry basis or phenomenological grounds \cite{JACKIW}. We list some model applications of IREG. Chiral gauge theories were discussed within IREG in \cite{VIGLIONI,JOILSON,BRUQUE}. The consistency conditions which led to a constrained and gauge invariant version of IREG in abelian theories appeared firstly in \cite{PRIM2} and were further discussed in \cite{PRD2012,ALEXANDRE,PRD1,gaugen}. In specific model calculations, IREG has been shown to respect supersymmetry in the Wess-Zumino model in \cite{JHEP} as well as in supersymmetric gauge theories \cite{eloy,IREGSUSY,HELVECIO} helping to shed light on the puzzle about IR contributions to the beta function of $N=1$ Super Yang-Mills theory. Because IREG parametrises regularization dependent parameters in perturbative calculations, it was useful to study some effective models of low energy QCD \cite{CO,DIAS}. For a study on the radiative origin of some Lorentz and CPT ambiguous terms in extended QED  within IREG we refer to \cite{PRD2,CPTEPJC,GAZZOLACPT}. In \cite{PRD3} we compared for the first time IREG to the BPHZ and dimesional methods and in \cite{RIRD} it was shown that the rules of IREG and DIFR can be made equivalent, at least to one loop order. The role played by quadratic divergences on phenomenology could de clearly discussed in the framework of IREG \cite{QUADRATIC}. Some applications of such discussion can be found in \cite{JEAN,JEANTOMS, JEAN2}. In \cite{GRAFENO} we have a field theoretical derivation  of interaction corrections to the universal  conductivity of graphene using perturbative tools based on IREG showing explicitly the origin of discrepancies in previous calculations using different regularisations. The present contribution is a step forward a generalisation of IREG beyond one loop order to non-abelian gauge theories. We show in a set of examples that just as in the abelian case a constrained version of IREG automatically deliver gauge invariant amplitudes. For this purpose it is necessary to closely compare IREG with fully dimensional schemes (CDR) and mixed schemes such as DRED.

It is noteworthy that even for methods that  work in the physical dimension,  the renormalisation of divergent Feynman amplitudes can lead to  inconsistencies in the manipulation of Dirac algebra in odd dimensions and  $\gamma_5$ matrix algebra  just as in dimensional methods. The crux of the matter is simple enough. By claiming the validity of the following properties:  (a) shift invariance (namely a property related to translational and momentum routing invariance in Feynman diagrams) , (b) linearity of  renormalisation (subtraction of the divergent content within a certain scheme) and (c) numerator/denominator consistency (such as for $k$ being a loop quadri-momentum and $m$ a mass $(k^2+m^2)/(k^2+m^2)=1$ in the integrand of a Feynman amplitude), the contraction of Lorentz indices does not commute with renormalisation. This is  the reason of some spurious anomalies in chiral gauge theories \cite{VIGLIONI, JOILSON, BRUQUE}. Because such properties are fundamental to comply with perturbative proofs of quantum action principles,  some operations in the physical dimension become forbidden in IREG (for instance, symmetric integration in a divergent integral). In \cite{BRUQUE} it was discussed that for the purpose of a consistent treatment of the $\gamma_5$-algebra in Feynman amplitudes, forbidding such operations in non-dimensional regularisations may not be sufficient. Consistency can be re-established formally by  avoiding some $n$-dimensional relations before renormalisation. The make such statement more rigorous one can proceed  analogously to  some consistent generalisations of DRED such as FDH, but in a much simpler fashion. The inherent inconsistencies found in DRED by Siegel \cite{SIEGEL} were solved by forbidding the use of certain $n$-dimensional relations before renormalisation \cite{AVDEEV}. In this construction the $n$-dimensional space to be used in DRED is not the genuine $n$-dimensional space (nS), but a quasi-$n$-dimensional space (QnS) as schematised in table \ref{tab:DIM}. Its relation with QdS in CDR is given by the direct-sum structure $QnS = QdS \oplus Q \varepsilon S$. In a similar fashion, one could define
IREG in a $QnS$ space with the difference that one needs not embed a $QdS$ space as in DRED or FDH. Formally we would have
$$
QnS|_{IREG}= nS \oplus X,
$$
where $nS$ is the genuine $n$-dimensional space which allows for using standard identities valid in $nS$ spaces at some steps of the calculations. Another advantage is that contrarily to the 
$Q \varepsilon S$ space, the $X$-space is only formally defined and does not require further explicit integration rules. In this way, the $\gamma_{5}$ will pertain to $nS$, and genuine dimensional identities cannot be applied, in general. This formal extension has however a few drawbacks \cite{BRUQUE} (which are also present in dimensional methods) and need further study: the generalisation of standard Dirac algebra to odd dimensions cannot be preserved and there is no finite complete set in Dirac space and hence the standard Fierz identities do not hold.

We emphasise that, in the present contribution, chiral theories will not be considered. Thus, we will not need to consider this space structure any further since, from a practical point of view, no distinction among them can be drawn in the examples to be studied. This fact allows us to work in a strictly four-dimensional framework whose rules we present in the next section. Nonetheless it is worth mentioning that the rules of IReg to handle chiral theories, so far in selected examples, do not require to abandon the physical dimension and are corroborated by an analysis in the space structure mentioned.
%After this brief review of the status of IREG, we present in the next section the defining rules of the method.

\subsubsection{The rules of IREG}

We start by introducing some terminology. We denote a general $n$-loop Feynman amplitude as ${\cal A}_n$ with $L$ external legs, $k_l$ being the internal (loop) momenta ($l= 1\cdots n$) and $p_i$ the external momenta. The divergent content of the amplitude will separated in a set of basic divergent integrals (BDI) in such a way that:
i) each overall-divergent amplitude is separated into a unique finite expression plus a divergent part, ii) power-counting finite expressions are not modified, and iii) linearity under the regularisation operation $R$ is preserved namely, $[aF + b G]^{R} = a[F]^{R}+b[G]^{R}$, where $F$ and $G$ are Feynman integrals, $a,\;b$ are constants (dependent on external momenta and/or masses). Moreover, invariance under shifts of the integration momenta and numerator-denominator consistency will be required. These two conditions are necessary in perturbative proofs of the quantum action principle \cite{QAP}. 
In order to fulfill all these properties, a definition of normal form\footnote{By normal form it is meant that by following the rules of the method, one arrives at unequivocal expressions.} in the context IREG is required, which renders consistency to the method. In this contribution we extend the analysis presented in \cite{BRUQUE}, by proposing a definition of normal form to multi-loop integrals in a way consistent with gauge invariance as our results show.

Given the general integral ${\cal A}_n$, we propose that a normal form is achieved after the two steps:

%Our results show that the above mentioned conditions are sufficiently stringent and general to allow for the abiding normal form to be applicable in gauge invariant perturbative multi-loop calculations in IREG, systematised as follows: 

\begin{enumerate}[label=\Alph*]
\item Perform the internal symmetry group and the usual Dirac algebra. As extensively discussed in \cite{BRUQUE}, identities only valid in $nS$ such as $\{\gamma_5,\gamma_\mu\}=0$ must not be used inside divergent amplitudes \cite{BRUQUE,VIGLIONI,JOILSON}. 

\item 
%\risk{Perform all available simplifications regarding metric contractions coming from derivative vertexes and/or dirac algebra.} 
The requirement of numerator/denominator consistency implies that terms with internal momenta squared in the numerator must be canceled against denominator. 
For instance,
\begin{align}
\int_{k,q}\frac{k^2}{k^2 q^2 (k-q)^2}\bigg|_{\text{IREG}} = \int_{k,q}\frac{1}{q^2(k-q)^2}\bigg|_{\text{IREG}},
\end{align}
where $\int_k \equiv \int d^4k/(2 \pi)^4$. 
% Notice, however, that one may not manipulate the numerator to enforce cancellations, since this may modify the normal form.  For instance,
% \begin{align}
% \int_{k,q}\frac{k.q}{k^2 (k-p)^2 q^2 (k-q)^2}\bigg|_{\text{IREG}}\neq \int_{k,q}\frac{k^2 + q^2 - (k-q)^2}{2 k^2 (k-p)^2 q^2 (k-q)^2}\bigg|_{\text{IREG}}.
% \end{align}
% \AC{I THINK WE SHOULD INDICATE HERE WHY THIS IS NOT SO, PROOF ALONG THE LINES OF THE APPENDIX ? OR JUST SAY HOW THIS EQUALITY SPOILS OUR PREMISES: SHIFT? NUMERATOR DENOMINATOR CONSISTENCY, CONTRACTION-RENORMALISATION NON COMMUTATIVITY?}
In the same vein, symmetric integration in divergent amplitudes cannot be enforced. That is, 
	\beq
	\Bigg[\int_k k^{\mu_1}\cdots k^{\mu_{2m}} f(k^2)\Bigg]^{\text{IREG}}\neq\;\frac{g^{\{\mu_1 \mu_2 } \cdots g^{\mu_{2m-1} \mu_{2m} \}}}{(2m)!} \Bigg[\int_k k^{2m} f(k^2)\Bigg]^{\text{IREG}},
	\label{eq:sym}
	\eeq
where  the curly brackets here indicate symmetrisation over Lorentz indices.
\end{enumerate}
After these steps, the resulting multi-loop integrand can  be manipulated consistently in the framework of IREG, meaning that i) each overall-divergent integral is separated into a unique finite expression plus a divergent part. 
Moreover, the UV content of ${\cal A}_n$ can be cast in terms of well-defined basic divergent integrals, which need not to be evaluated. We assume  without loss of generality that all the masses of the underlying model are zero in order to define a massless minimal subtraction scheme. 

Given the normal form defined by the steps A and B above, in the following we present the rules of IREG for  UV divergent amplitudes necessary to evaluate $\beta$-functions of gauge couplings in abelian and non-abelian theories. For the treatment of both UV and IR divergences in IREG including factorisation properties of  Feynman amplitudes, a more proficuous arena is S-matrix calculations involving cross-sections and decay rates to be discussed elsewhere.

\begin{enumerate}

	\item Starting at one loop, assume an implicit regulator, say a momentum cut-off, in order to remove external momenta dependence from the divergent part of the  amplitude by judiciously applying the identity
	\begin{align}
	\frac{1}{(k_{l}-p_{i})^2-\mu^2}=\sum_{j=0}^{n_{i}^{(k_{l})}-1}\frac{(-1)^{j}(p_{i}^2-2p_{i} \cdot k_{l})^{j}}{(k_{l}^2-\mu^2)^{j+1}}
	+\frac{(-1)^{n_{i}^{(k_{l})}}(p_{i}^2-2p_{i} \cdot k_{l})^{n_{i}^{(k_{l})}}}{(k_{l}^2-\mu^2)^{n_{i}^{(k_{l})}}
		\left[(k_{l}-p_{i})^2-\mu^2\right]},
	\label{ident}
	\end{align}
	in the propagators. Here $\mu \downarrow 0$ is a fictitious mass (infrared regulator) and at one loop order $k_l$ is simply $k$. It should be emphasized that, since the starting integrals are IR-safe, the infrared regulator will only be needed in intermediate steps of the calculation, canceling in the end result. Therefore, gauge invariance will not be spoiled. Basic divergent integrals (BDI's) appear as
	\bea
	I_{log}(\mu^2)&\equiv& \int_{k} \frac{1}{(k^2-\mu^2)^{2}},\quad \quad
    I_{log}^{\nu_{1} \cdots \nu_{2r}}(\mu^2)\equiv \int_k \frac{k^{\nu_1}\cdots
		k^{\nu_{2r}}}{(k^2-\mu^2)^{r+2}}, \nonumber \\
    I_{quad}(\mu^2)&\equiv& \int_k \frac{1}{(k^2-\mu^2)},\quad \quad
	I_{quad}^{\nu_{1} \cdots \nu_{2r}}(\mu^2)\equiv \int_k \frac{k^{\nu_1}\cdots
		k^{\nu_{2r}}}{(k^2-\mu^2)^{r+1}}.
	\eea
	The UV finite part in the limit where $\mu \downarrow 0$ has logarithmical dependence in the physical momenta which is the characteristic behaviour of the finite part of massless amplitudes \cite{LOGS}.

\item BDI's with Lorentz indices $\nu_{1} \cdots \nu_{2r}$ are systematically reduced to linear combinations of BDI's without Lorentz indices (with the same superficial degree of divergence) since we comply with invariance under shifts of the integration momenta and numerator-denominator consistency \cite{BRUQUE}. Therefore, total derivatives with respect to the internal momenta must vanish, e.g.
\bea
\int_k\frac{\partial}{\partial k_{\mu}}\frac{k^{\nu}}{(k^{2}-\mu^{2})^{2}}&=&4\Bigg[\frac{g_{\mu\nu}}{4}I_{log}(\mu^2)-I_{log}^{\mu\nu}(\mu^2)\Bigg]=0,\label{ST1L}
\\
\int_k\frac{\partial}{\partial k_{\mu}}\frac{k^{\nu}}{(k^{2}-\mu^{2})}&=& 2\Bigg[\frac{g_{\mu\nu}}{2}I_{quad}(\mu^2)-I_{quad}^{\mu\nu}(\mu^2)\Bigg]=0.
\label{ST1Q}
\eea

\item An arbitrary positive (renormalisation group) mass scale $\lambda$
appears via  regularisation independent identities, for instance
\beq
I_{log}(\mu^2) = I_{log}(\lambda^2) + \frac{i}{(4 \pi)^2} \ln \frac{\lambda^2}{\mu^2},
\label{SR1}
\eeq
which enables us to write a BDI as a function of $\lambda^2$ plus logarithmic functions of $\mu^2/\lambda^2$, $\mu$ being a fictitious mass which is added to massless propagators. $I_{quad} (\mu^2)$ can be chosen to vanish as $\mu$ goes to zero as we will show through a general parametrisation of BDI's. The limit
$\mu \rightarrow 0$ is well defined for the whole amplitude  since it
is power counting infrared convergent ab initio. The BDI can be absorbed in the renormalisation constants (without explicit evaluation) \cite{CLANT} and renormalisation functions can be computed using the regularisation independent identity:
\beq
\lambda^2\frac{\partial I_{log}(\lambda^2)}{\partial \lambda^2}= -\frac{i}{(4 \pi)^2}.
\eeq

\item At higher loop order the divergent content can be expressed in terms of  BDI in one loop momentum after performing $n-1$ integrations. The order of such integrations is chosen systematically to display the counterterms to be subtracted in compliance with the Bogoliubov's recursion formula \cite{BPHZ,ADRIANO}. The general form of the terms of a Feynman amplitude after $l$ integrations is
\begin{align}
&I^{\nu_{1}\ldots \nu_{m}}\!=\!\!\int\limits_{k_{l}}\!\frac{A^{\nu_{1}\ldots \nu_{m}}(k_{l},q_{i})}{\prod_{i}[(k_{l}-q_{i})^{2}-\mu^{2}]}\ln^{l-1}\!\left(\!-\frac{k_{l}^{2}-\mu^{2}}{\lambda^{2}}\right)\!,
\label{I}
\end{align}
\noindent
where $l=1, \cdots , n$ and $q_{i}$ is an element (or combination of elements) of the set $\{p_{1},\ldots,p_{L},k_{l+1},\ldots,k_{n}\}$.   $A^{\nu_{1}\ldots \nu_{m}}(k_{l},q_{i})$ represents all possible combinations of $k_{l}$ and $q_{i}$ compatible with the Lorentz structure.
\item Apply relation (\ref{ident}) in (\ref{I}) by choosing $n_{i}^{(k_{l})}$ such that all divergent integrals are free of $q_{i}$. Therefore the divergent integrals are cast as a combination of
\begin{align}
I_{log}^{(l)}(\mu^2)&\equiv \int\limits_{k_{l}} \frac{1}{(k_{l}^2-\mu^2)^{2}}
\ln^{l-1}{\left(-\frac{k_{l}^2-\mu^2}{\lambda^2}\right)},\quad
\label{Ilogilog}\\
I_{log}^{(l)\nu_{1} \cdots \nu_{2r}}(\mu^2)&\equiv \int\limits_{k_{l}} \frac{k_{l}^{\nu_1}\cdots
	k_{l}^{\nu_{2r}}}{(k_{l}^2-\mu^2)^{r+1}}
\ln^{l-1}{\left(-\frac{k_{l}^2-\mu^2}{\lambda^2}\right)},
\label{IlogLorentz}\\
I_{quad}^{(l)}(\mu^2)&\equiv \int\limits_{k_{l}} \frac{1}{(k_{l}^2-\mu^2)}
\ln^{l-1}{\left(-\frac{k_{l}^2-\mu^2}{\lambda^2}\right)},\quad
\label{Iquadiquad}\\
I_{quad}^{(l)\nu_{1} \cdots \nu_{r+2}}(\mu^2)&\equiv \int\limits_{k_{l}} \frac{k_{l}^{\nu_1}\cdots
	k_{l}^{\nu_{2r}}}{(k_{l}^2-\mu^2)^{r+1}}
\ln^{l-1}{\left(-\frac{k_{l}^2-\mu^2}{\lambda^2}\right)}.
\label{IquadLorentz}
\end{align}
As before, higher loop BDI's are reduced to scalar ones by considering vanishing total derivatives

\begin{align}
\int_k\frac{\partial}{\partial k_{\nu_{1}}}\frac{k^{\nu_{2}}\cdots k^{\nu_{2j}}}{(k^{2}-\mu^{2})^{1+j-i}}\ln^{l-1}\Bigg[-\frac{(k^{2}-\mu^{2})}{\lambda^{2}}\Bigg]=0.
\label{tsdef}
\end{align}
For instance,
\begin{align}
&I_{log}^{(l)\,\mu
	\nu}(\mu^2)=\sum_{j=1}^{l}\left(\frac{1}{2}\right)^j\!\frac{(l-1)!}{(l-j)!}\!\left\{\frac{g^{\mu \nu}}{2}I_{log}^{(l-j+1)}(\mu^2)\right\}.
\label{identsurface1}
\end{align}
\item A renormalisation group scale is encoded in BDI's. At n$^{th}$-loop order a relation analogous to (\ref{SR1}) is obtained via the regularisation independent identity
\begin{align}
I_{log}^{(l)}(\mu^2)&=I_{log}^{(l)}(\lambda^2)-\frac{b}{l}\ln^{l}\left(\frac{\mu^2}{\lambda^2}\right)- b \sum_{j=1}^{l-1}\frac{(l-1)!}{(l-j)!}\ln^{l-j}\left(\frac{\mu^2}{\lambda^2}\right),
\label{scale}\\
\mbox{where}\quad\lambda^{2}&\neq0,\;\;  b \equiv\frac{i}{(4\pi)^2}.
\label{bd}
\end{align}
\item BDI's can be absorbed in renormalisation constants. A minimal, mass-independent scheme amounts to absorb only $I_{log}^{(l)}(\lambda^2)$. To evaluate RG constants, BDI's need not be explicitly evaluated as their derivatives with respect to the renormalisation scale $\lambda^2$ are also BDI's. For example \cite{PRD2012},
\begin{align}
\lambda \frac{\partial I_{log}(\lambda^2)}{\partial \lambda^2} &= -b,\quad
 \lambda^2 \frac{\partial I_{quad} (\lambda^2)}{\partial \lambda^{2}}= \lambda^2 I_{log}(\lambda^2),\nonumber \\
\lambda^2\frac{\partial I_{log}^{(n)}(\lambda^2)}{\partial \lambda^{2}}&=-(n-1)\, I_{log}^{(n-1)}(\lambda^2)- b \,\, \alpha^{(n)}\, ,\nonumber \\
\lambda^2\frac{\partial I_{log}^{(n)\,\mu\nu}(\lambda^2)}{\partial \lambda^{2}}&=-(n-1)I_{log}^{(n-1)\,\mu\nu}(\lambda^2)-\frac{g_{\mu\nu}}{2} \, b \, \Upsilon^{(n)}.
\label{gerder}
\end{align}
where $n \ge 2$, $\alpha^{(n)} = (n-1) !$ and $\Upsilon^{(n)}$ may be obtained from $\alpha^{(n)}$ via relation (\ref{identsurface1}).

\end{enumerate}

Finally, some comments are in order. For simplicity, we will discard terms quadratically divergent encoded as $I_{quad}^{(l)}(\mu^2)$ since they must cancel in theories that are multiplicative renormalizable, which are the ones we consider here. Moreover, in the framework of IREG, one is not allowed, in general, to evaluate a sub-diagram and join the obtained result in the full diagram. The reason can be traced back to equations similar to (\ref{eq:sym}). This fact does not amount in a violation of unitarity since, according to the BPHZ theorem, only the divergent content of the integrals in (\ref{eq:sym}) must coincide, and they do. However, local terms may be generated which could (possibly) violate gauge invariance. Therefore, we argue that the normal form obtained by the steps (A) - (B) is the one that respects both unitarity and gauge invariance. 

\begin{figure}[h!]
\centering
\includegraphics[width=0.20\textwidth]{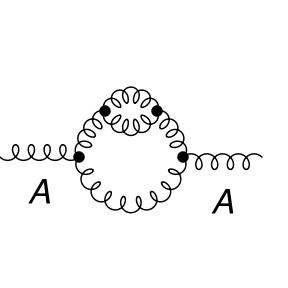}%
\caption{Two-loop diagram which contains a gluon loop as sub-diagram}
\label{fig:Feynman}
\end{figure}

In order to clarify the discussion we present below an example. Consider the diagram of Fig. \ref{fig:Feynman}, 
\noindent
whose amplitude (in the Feynman gauge) is schematically given by
\begin{equation}
\mathcal{A} \propto \int_k \Pi_{\mu\nu\alpha\beta}(k,p)\int_l \frac{\mathcal{F}^{\alpha\beta}(l,k,p)}{l^2(l-k)^2},
\end{equation}
where $l$ is the internal momentum of the sub diagram (gluon loop), $k$ the internal momentum of the complete diagram, and $p$ the external momentum. From the many possibilities encoded in $\mathcal{F}^{\alpha\beta}$, one is simply $l^{\alpha}l^{\beta}$ which, after contraction with $g_{\alpha\beta}k_{\mu}k_{\nu}$, one of the many terms presented in $\Pi_{\mu\nu\alpha\beta}(k,p)$, generates a contribution as
\begin{align}
\mathcal{B} \propto &\int_k \frac{k_{\mu}k_{\nu}}{k^4(k-p)^2}\int_l \frac{l^2}{l^2(l-k)^2}\label{eq:b1}\\
\propto &\int_k \frac{k_{\mu}k_{\nu}}{k^4(k-p)^2}\int_l \frac{1}{(l-k)^2}\label{eq:b2}
\end{align}
Notice that eq.(\ref{eq:b1}) was obtained by applying the step (A) of our procedure, while eq.(\ref{eq:b2}) is due to step (B). Next one should evaluate the integrals in IREG according to the algorithm presented in \cite{ADRIANO}, and the rules sketched before.
% in the last subsection. 
It is not difficult to perceive that a null result will be obtained, since we are allowed to perform shifts and the integral in $l$ will be identified with $I_{quad}(\mu^2)$ which we drop as already explained. On the other hand, if one opts to evaluate the sub-diagram independently, one obtains 
\begin{align}
\int_l \frac{l^{\alpha}l^{\beta}}{l^2(l-k)^2}\Bigg|_{\text{IREG}}=k^{\alpha}k^{\beta}\mathcal{T}_{1}+g^{\alpha\beta}k^{2}\mathcal{T}_{2},
\end{align}
where $\mathcal{T}_{1}$ and $\mathcal{T}_{2}$ are scalar functions defined in eq. \ref{app:1loop}. By contracting the previous equation with $g_{\alpha\beta}k_{\mu}k_{\nu}$ one now gets
\begin{align}
\mathcal{\tilde{B}} \propto & \int_k \left[\frac{k_{\mu}k_{\nu}k^2}{k^4(k-p)^2}(\mathcal{T}_{1}+4\mathcal{T}_{2})\right]=\int_k \left[\frac{k_{\mu}k_{\nu}k^{2}}{k^4(k-p)^2}\left(-\frac{b}{6}\right)\right]
\end{align}
which is (clearly) different from zero.

%%%%%%%%%%%%%%%%%%%%%%%%%%%%%%%%%%%%%%%
\subsection{Correspondence among IREG and dimensional methods}
\label{sec:correlation} 
%%%%%%%%%%%%%%%%%%%%%%%%%%%%%%%%%%%%%%%

In this section we analyse to which extent it is possible to recover results for amplitudes evaluated by dimensional methods once the result in IREG is known. In IREG, the  UV log-divergent content of a Feynman amplitude is expressed by the BDI's 
$I_{log}^{(n)}(\lambda^2)$, while in dimensional methods they appear as poles in $\epsilon\rightarrow 0$.  Therefore, given a n-loop amplitude, one may wonder if extracting the residues of $\epsilon^{-n}$  by evaluating $I_{log}^{(n)}(\lambda^2)$ in $4 - 2 \epsilon$ dimensions is sufficient to map the IREG result on the one obtained in CDR or DRED, for instance. Starting at $n=1$, consider the log-divergent integral composed by the product of two massless propagators (scalar self-energy amplitude): 
\begin{equation}
\mathcal{I}=\int_{k}\frac{1}{k^2(k-p)^2} \stackrel{\text{IREG}}{=} I_{log}(\lambda^{2}) - b\ln\left[-\frac{p^2}{\lambda^{2}}\right] + 2b.
\label{eq:self}
\end{equation} 
By evaluating $I_{log}(\lambda^2)$ in $d=4 - 2 \epsilon$ dimensions and expanding around $\epsilon=0$ \cite{ZURICH}
\begin{equation}
I_{log}^d(\lambda^2)= (\mu_{DR})^{4-d}\int\frac{d^{d}k}{(2\pi)^{d}}\frac{1}{(k^{2}-\lambda^{2})^{2}} = b \left[\frac{1}{\epsilon} - \gamma_{E} + \ln(4\pi) +\ln\left(\frac{\mu^{2}_{DR}}{\lambda^{2}}\right)\right],
\label{eq:log dreg0}
\end{equation}
one obtains
\begin{align}
\mathcal{I^{\text{IREG}}}\Big|_d = b \left[\frac{1}{\epsilon} - \gamma_{E} + \ln(4\pi) - \ln\left[-\frac{p^2}{\mu^{2}_{DR}}\right] + 2\right].
\label{eq:id ireg}
\end{align}
Notice that the $\lambda^2$ dependence is automatically traded by $\mu^{2}_{DR}$.
This result coincides with CDR and DRED, since contractions of the metric are absent, namely
\begin{align}
\mathcal{I}_d=b\left[\frac{1}{(4\pi)^{-\epsilon}}\left(-\frac{\mu^{2}_{DR}}{p^2}\right)^{\epsilon}
\frac{\Gamma(\epsilon)\Gamma^{2}(1-\epsilon)}{\Gamma(2-2\epsilon)}\right]= \mathcal{I^{\text{IREG}}}\Big|_d.
\label{eq:dred}
\end{align}
Therefore, for this one-loop example, evaluating BDI's in $4 - 2 \epsilon$ dimensions is equivalent to dimensional methods. It should be pointed out that at one-loop order, when DRED and CDR yield different finite parts, we expect to recover the results of the former. The reason is simply IREG does not recourse to dimensional continuation and thus contractions such as $g_{ab}g^{ab}$ give 4 rather than $d=4-2\epsilon$. Moreover, at least at one-loop order, by identifying the renormalisation scales of IREG and dimensional methods in eq. \ref{eq:log dreg0}, one notices that the subtraction of BDI's  is equivalent to the $\overline{\text{MS}}$ scheme. Interestingly, in FDR scheme \cite{ZURICH,FDR}, $\mu^{2}$ is replaced by $\mu^{2}_{DR}$ in the final result (after taking the limit $\mu\downarrow 0$), in order to reproduce the $\overline{\text{MS}}$ scheme. As for quadratic divergences a similar approach yields
\begin{equation}
I_{quad}(\mu^2)= (\mu_{DR})^{4-d}\int\frac{d^{d}k}{(2\pi)^{d}}\frac{1}{(k^{2}-\mu^{2})} = b \mu^{2}\left[1+\frac{1}{\epsilon} +\ln\left(\frac{\mu^{2}_{DR}}{\mu^{2}}\right)\right],
\end{equation}
which clearly vanishes in the limit $\mu \downarrow 0$. From the point of view of IREG, one could still keep quadratic divergences as a BDI, which cancel out in multiplicatively renormalisable theories \cite{ELOY,PRD2013,QUADRATIC} and thus they can be dismissed in massless models. 

We proceed to study the correspondence between dimensional methods and IREG to higher loop order where some  subtleties appear. In analogy to the self-energy amplitude we studied at one-loop, consider the n-loop product of massless bubble diagrams proportional to  the integrals:
\begin{align}
\mathcal{J}=\int_{k_{1}}\frac{1}{k_{1}^2(k_{1}-p)^2}\cdots\int_{k_{n}}\frac{1}{k_{n}^2(k_{n}-p)^2}.
\label{eq:disjoint}
\end{align} 
Using the rules of IREG, each integral in $k_{i}$ can be independently  performed to give
% \begin{align}
% \mathcal{J}=\left[I_{log}(\lambda^{2}) - b\ln\left[-\frac{p^2}{\lambda^{2}}\right] + 2b\right]^{n}\xrightarrow[\text{DRED,CDR}]{\text{?}}b^{n}\left[\frac{1}{\epsilon}- \gamma_{E} + \ln(4\pi)- \ln\left(-\frac{p^2}{\mu_{DR}^{2}}\right) + 2\right]^{n},
% \label{eq:buble}
% \end{align}
\begin{align}
\mathcal{J^{\text{IREG}}}&=\left[I_{log}(\lambda^{2}) - b\ln\left[-\frac{p^2}{\lambda^{2}}\right] + 2b\right]^{n}.
\end{align}
In order to establish a correspondence with dimensional methods at n-loop order, the use of eq. \ref{eq:id ireg} is inappropriate, since terms $\mathcal{O}(\epsilon)$ are needed. 
% obtain
% \begin{align}
% \mathcal{J^{\text{IREG}}}\Big|_d&=\left(\mathcal{I^{\text{IREG}}}\Big|_d\right)^{n}=b^{n}\left[\frac{1}{\epsilon}- \gamma_{E} + \ln(4\pi)- \ln\left(-\frac{p^2}{\mu_{DR}^{2}}\right) + 2\right]^{n}.
% \label{eq:buble}
% \end{align}
% %where to establish the correspondence we have used eq. \ref{eq:id ireg} and since no Lorentz indices contractions appear we expect the same result within CDR or DRED.  
% However, , it is immediate to see, that at higher loop order, it is necessary to include
%
% we can verify through this example that only some of the coefficients of the poles in $\epsilon$ will be identical to dimensional methods should we evaluate a BDI in $d=4-2\epsilon$ dimensions in the end of the calculation. To see that, 
%consider 
Therefore, one should consider $I_{log}(\lambda^{2})$ before expanding around $\epsilon=0$,
\begin{equation}
I_{log}^{d}(\lambda^2)= (\mu_{DR})^{4-d}\int\frac{d^{d}k}{(2\pi)^{d}}\frac{1}{(k^{2}-\lambda^{2})^{2}} = \frac{b}{(4\pi)^{-\epsilon}} \left(\frac{\mu^{2}_{DR}}{\lambda^{2}}\right)^{\epsilon}\Gamma(\epsilon),
\label{eq:log dreg}
\end{equation}
and let us rewrite $\ln\left(-p^2/\lambda^{2}\right)$ so that the scale $\mu^{2}_{DR}$ emerges:
\begin{equation}
\ln\left(-\frac{p^2}{\lambda^{2}}\right)= \lim_{\epsilon\downarrow 0}
\frac{\Gamma(\epsilon)}{(4\pi)^{-\epsilon}}\left[\left(\frac{\mu_{DR}^{2}}{\lambda^{2}}\right)^{\epsilon}-\left(-\frac{\mu^{2}_{DR}}{p^2}\right)^{\epsilon}\right],
\label{eq:log}
\end{equation}
where we have used that $\ln (x) = \lim_{\epsilon\downarrow 0}\,\,[(x^\epsilon-1)/\epsilon + \alpha(\epsilon)]$, $\alpha(\epsilon\downarrow 0)=0$, and to write in terms of the gamma function  we have chosen  $\alpha (\epsilon)= (\ln 4\pi - \gamma_E)[(\mu_{DR}^2/\lambda^2)^\epsilon-(-\mu_{DR}^2/p^2)^\epsilon] $. Thus,
% \begin{align}
% \mathcal{I}=\left[I_{log}(\lambda^{2}) - b\ln\left[-\frac{p^2}{\lambda^{2}}\right] + 2b\right]^{n}\xrightarrow[\text{DRED,CDR}]{\text{?}}b^{n}\left[\frac{1}{(4\pi)^{-\epsilon}} \left(-\frac{\mu^{2}_{DR}}{p^2}\right)^{\epsilon}\Gamma(\epsilon) + 2\right]^{n},
% \label{eq:buble new}
% \end{align}
\begin{align}
\mathcal{J^{\text{IREG}}}\Big|_d=b^{n}\left[\frac{1}{(4\pi)^{-\epsilon}} \left(-\frac{\mu^{2}_{DR}}{p^2}\right)^{\epsilon}\Gamma(\epsilon) + 2\right]^{n},
\label{eq:buble new}
\end{align}
which, by construction, is independent of $\lambda$.
On the other hand, by performing the computation in DRED from the start one obtains
\begin{align}
\mathcal{J}_{d}&=b^{n}\left[\frac{1}{(4\pi)^{-\epsilon}}\left(-\frac{\mu^{2}_{DR}}{p^2}\right)^{\epsilon}
\frac{\Gamma(\epsilon)\Gamma^{2}(1-\epsilon)}{\Gamma(2-2\epsilon)}\right]^{n}.
\label{eq:dred new}
\end{align}
Now it can be seen that eqs. \ref{eq:buble new} and \ref{eq:dred new} only agree in the $\epsilon^{-n}$ and $\epsilon^{-n+1}$ coefficients. This implies that, in a two-loop computation, the pole structure of an integral of the type of eq. \ref{eq:disjoint} can be fully recovered from the IREG result. Notice that this may not be the case for a CDR computation if, as already pointed out, contractions of the metric are present. For instance, consider the two-loop integral 
\begin{align}
\mathcal{J}&=\int_{k_{1},k{2}}\frac{(k_{1}.k_{2})^{2}}{k_{1}^2(k_{1}-p)^2k_{2}^2(k_{2}-p)^2}=\int_{k_{1}}\frac{(k_{1})^{\alpha}(k_{1})^{\beta}}{k_{1}^2(k_{1}-p)^2}\int_{k{2}}\frac{(k_{2})_{\alpha}(k_{2})_{\beta}}{k_{2}^2(k_{2}-p)^2}\nonumber\\
&=\left[g^{\alpha\beta}p^{2}A+p^{\alpha}p^{\beta}B\right]\left[g_{\alpha\beta}p^{2}A+p_{\alpha}p_{\beta}B\right]=\left[\textbf{4}A^{2}+2p^{2}AB+p^{4}B^{2}\right]
\end{align} 
where, due to a contraction of the type $g_{ab}g^{ab}=\textbf{4}$ we can only recover the divergent results of DRED, not CDR. Therefore, by considering a multi-loop amplitude containing only disjoint one-loop integrals as eq. \ref{eq:disjoint}, we can already draw some conclusions: in general, given the IREG result of a multi-loop amplitude, it is not possible to recover the result from DRED by just evaluating BDI's in $4-2\epsilon$ dimensions. For specific cases, like eq. \ref{eq:disjoint} at two-loop order, one can recover all divergent terms ( $\mathcal{O}(\epsilon^{-2})$ and $\mathcal{O}(\epsilon^{-1})$), but not the finite part of the amplitude. The mismatch in the finite part results from the inclusion of $O(\epsilon)$ terms in the starting expression for the comparison of IREG and DRED in \ref{eq:log}. These terms are irrelevant at 1-loop order but generate in the sub-leading orders of  higher loops finite terms from cross products of 1/$\epsilon$ powers and $\mathcal{O}(\epsilon)$ terms. One should emphasize that therefore this mismatch  is neither a shortcoming of dimensional  methods nor of IREG, but a simple artifact resulting from the bridging of two approaches at one-loop order. Nevertheless, for this kind of amplitude, one can always retrieve the $\epsilon^{-n}$ and $\epsilon^{-n+1}$ terms. This finding may be useful to check intermediate steps in a computation done with IREG by comparing with its counterpart in DRED, for instance. Another conclusion to be drawn is that possibly the subtraction scheme in IREG given by the removal of BDI's such as $I_{ln}^{(n)}(\lambda^2)$ may not correspond to the $\overline{\text{MS}}$ scheme or even the $\overline{\text{DS}}$ scheme. However, since this conclusion can only be ascertained by the computation of renormalization constants rather than amplitudes, more investigations are necessary which we will perform elsewhere.

To conclude this section, we show that, in general, one can only expect to reproduce the $\mathcal{O}(\epsilon^{-n})$ term of a n-loop amplitude obtained within CDR or DRED by evaluating BDI's in $4-2\epsilon$ dimensions. This happens because in general one also has denominators of the form  $(k_{i}-k_{j})^{2}$. 
%
%
% Therefore, even in this simple case where all n-loop integrals are independent of each other, one can only expect to reproduce the $\epsilon^{-n}$ and $\epsilon^{-n+1}$ coefficients of DRED by evaluating BDI's in dimensional schemes. The situation is even more restrictive when there are denominators of the form  $(k_{i}-k_{j})^{2}$) as we are going to show. 
To illustrate this point, we consider a two-loop integral as below
\begin{align}
\mathcal{T}=\int_{k,l}\frac{1}{k^2(k-p)^2}\frac{1}{l^2(l-k)^2},
\end{align} 
where, given the rules of IREG, one must first evaluate the integral in $l$ 
\begin{align}
\mathcal{T}^{\text{IREG}}=\int_{k}\frac{1}{k^2(k-p)^2}\left[I_{log}(\lambda^{2}) - b\ln\left[-\frac{k^2}{\lambda^{2}}\right]+2b\right].
\end{align} 
As can be seen, due to the appearance of the denominator $(l-k)^{2}$, we have a non-local term on the internal momenta $k$, which will generate $I_{log}^{(2)}(\lambda^2)$. The UV divergent part of $\mathcal{T}$ is given by
\begin{align}
\mathcal{T}^{\text{IREG}}_{\text{div}}=I_{log}^{2}(\lambda^2)-bI_{log}^{(2)}(\lambda^2)-bI_{log}(\lambda^2)\ln\left[-\frac{p^2}{\lambda^{2}}\right]+4bI_{log}(\lambda^2).
\end{align} 

To obtain a correspondence with DRED, one could consider to replace powers of $I_{log}(\lambda^2)$ by eq. \ref{eq:log dreg}, as well as $\ln\left[-\frac{p^2}{\lambda^{2}}\right]$ by eq. \ref{eq:log}. Regarding $I_{log}^{(2)}(\lambda^2)$, one obtains
% \begin{align}
% I_{log}^{(2)}(\lambda^2)&= \int\frac{d^{4}k}{(2\pi)^{4}}\frac{1}{(k^{2}-\lambda^{2})^{2}} \ln\left[-\frac{k^2}{\lambda^{2}}\right]\nonumber\\
% &\xrightarrow[\text{DRED}]{\text{?}}(\mu^{2}_{DR})^{\epsilon}\int\frac{d^{d}k}{(2\pi)^{d}}\frac{1}{(k^{2}-\lambda^{2})^{2}}\frac{1}{(4\pi)^{-\epsilon}}\left[\left(\frac{\mu_{DR}^{2}}{\lambda^{2}}\right)^{\epsilon}-\left(-\frac{\mu^{2}_{DR}}{k^2}\right)^{\epsilon}\right]\Gamma(\epsilon)\nonumber\\
% &=\frac{b}{(4\pi)^{-2\epsilon}}\left(\frac{\mu_{DR}^{2}}{\lambda^{2}}\right)^{2\epsilon}\Gamma(\epsilon)\left[\Gamma(\epsilon)-\frac{\Gamma(2-4\epsilon)\Gamma(2\epsilon)}{\Gamma(2-3\epsilon)}\right],
% \label{eq:log2 dreg}
% \end{align}
\begin{align}
I_{log}^{(2)}(\lambda^2)&= \int\frac{d^{4}k}{(2\pi)^{4}}\frac{1}{(k^{2}-\lambda^{2})^{2}} \ln\left[-\frac{k^2}{\lambda^{2}}\right],\nonumber\\
I_{log}^{(2)}(\lambda^2)\Big|_{d}&=(\mu^{2}_{DR})^{\epsilon}\int\frac{d^{d}k}{(2\pi)^{d}}\frac{1}{(k^{2}-\lambda^{2})^{2}}\frac{1}{(4\pi)^{-\epsilon}}\left[\left(\frac{\mu_{DR}^{2}}{\lambda^{2}}\right)^{\epsilon}-\left(-\frac{\mu^{2}_{DR}}{k^2}\right)^{\epsilon}\right]\Gamma(\epsilon)\nonumber\\
&=\frac{b}{(4\pi)^{-2\epsilon}}\left(\frac{\mu_{DR}^{2}}{\lambda^{2}}\right)^{2\epsilon}\Gamma(\epsilon)\left[\Gamma(\epsilon)-\frac{\Gamma(2-4\epsilon)\Gamma(2\epsilon)}{\Gamma(2-3\epsilon)}\right],
\label{eq:log2 dreg}
\end{align}
where we have used eq. \ref{eq:log} on the second line. Considering only terms up to $\mathcal{O}(\epsilon^{-1})$ one thus obtains
\begin{align}
\mathcal{T}^{\text{IREG}}_{\text{div}}\Big|_{d}=\frac{b^{2}}{2\epsilon^{2}}+\frac{b^{2}}{\epsilon}\left[\frac{7}{2}-\gamma_{E}+\ln(4\pi)-\ln\left(-\frac{p^2}{\mu_{DR}^{2}}\right)\right].
\label{eq:I ireg}
\end{align}

If the calculation is done in DRED from the start, one gets
\begin{align}
\mathcal{T}\Big|_{d}=&\frac{b^2 (4 \pi )^{2 \epsilon } \Gamma (1-2 \epsilon ) \Gamma (1-\epsilon )^3 \Gamma (\epsilon ) \Gamma (2 \epsilon )
   \left(-\frac{p^2}{\mu ^2}\right)^{-2 \epsilon }}{\Gamma (2-3 \epsilon ) \Gamma (2-2 \epsilon ) \Gamma (\epsilon +1)}\nonumber\\
=&\frac{b^{2}}{2\epsilon^{2}}+\frac{b^{2}}{\epsilon}\left[\frac{5}{2}-\gamma_{E}+\ln(4\pi)-\ln\left(-\frac{p^2}{\mu_{DR}^{2}}\right)\right],
\end{align}
which differs from eq. \ref{eq:I ireg} in local terms of $\mathcal{O}(\epsilon^{-1})$. Nevertheless, the terms in $\epsilon^{-2}$ as well as non-local divergent terms can be reproduced. This observation allows us to check diagram-by-diagram the results we obtain in further sections. 

\section{IREG to two loop order: gauge theories}
\label{sec:results} 
%%%%%%%%%%%%%%%%%%%%%%%%%%%%%%%%%%%%%%%

In this section we apply the procedure discussed in the last section to a variety of examples. First, we discuss abelian theories such as scalar QED (which possesses derivative vertices) and spinorial QED (which requires a proper treatment of Dirac algebra \cite{BRUQUE}). Next, we turn to non-abelian gauge theories such as Yang-Mills, and finally QCD. We will be mainly interested in the computation of the two-loop coefficient of the gauge couplings $\beta$ function in all these theories. As is well-known, in the QED case one may only resort to the calculation of the two-point photon correction since $Z_{g}=Z_{A}^{-1/2}$ due to the Ward identity. As usual, $Z_{g/A}$ relates to the coupling/photon renormalisation function. In the case of non-abelian theories, the situation is more involved. In order to mimic the behavior of the QED case, one can resort to the background field method \cite{Abbott}, which guarantees the relation $Z_{g}=Z_{\hat{A}}^{-1/2}$ where $\hat{A}$ is now the gluon background field. This fact simplifies considerably the computation, since only two-point functions will be needed.

Therefore, in this contribution we will only deal with two-point functions with the photon (scalar/spinorial QED) or the gluon background field (Yang-Mills and QCD) as external legs. This implies that only topologies as below can appear 
\begin{figure}[h!]
\centering
\includegraphics[]{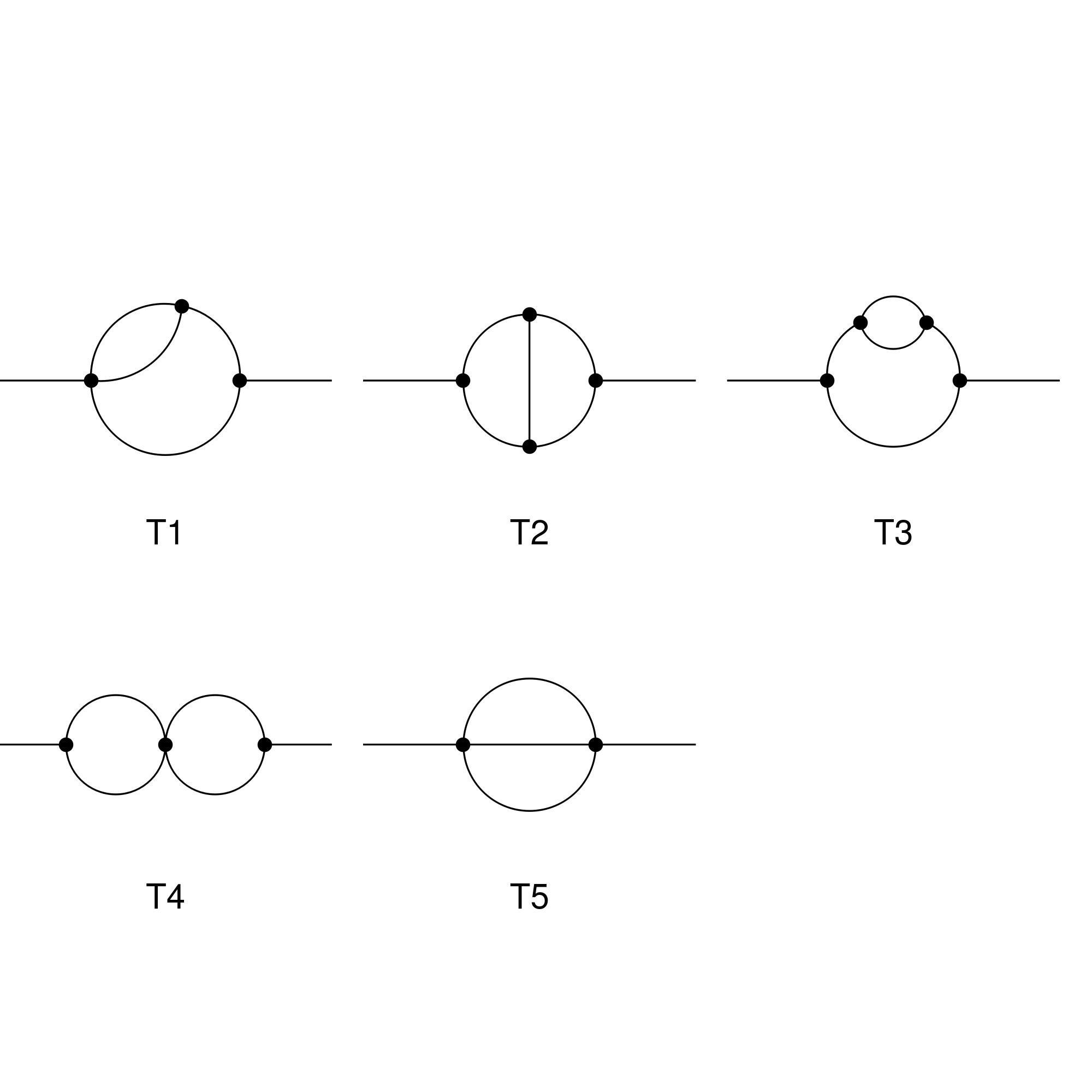}%
\caption{Two-loop topologies for two-point functions}
%\label{Feynman}
\end{figure}

\noindent
Notice that we are already omitting tapdole-like diagrams. The reason is twofold. Firstly, since we are only interested in the gauge coupling $\beta$ function, we can consider massless scalars/fermions. Secondly, given the minimal set of rules presented in the last section, we drop quadratically scaleless integrals, encoded as $I_{quad}^{(l)}(\mu^2)$. Given the general topologies, one has now to fill them with the field content of the theory at hand. For instance, in spinorial QED, only topologies T3 and T4 appear. In order to perform this task automatically, we have made use of \textit{FeynArts} \cite{FeynArts} which already has the spinorial QED model implemented. For the background field method, only the Electroweak Theory is already implemented, so we have adapted it to consider a background version of QCD as well. For scalar QED, we have opted to use the Feynman rules of \cite{Schwartz}, building the amplitudes ourselves. 

Once the amplitudes in all theories have been built, we adapted them to be recognized by \textit{FormCalc} \cite{FormCalc}, allowing \textbf{only} Dirac and Lorentz algebra to be performed with the end result for each topology given schematically by:

\begin{align}
\mathcal{A_{\text{T1}}} &\propto \int_{k,l} \frac{\mathcal{F}_{T1/T2}^{\alpha\beta}(l,k,p)}{k^2(k-p)^2 l^2(l-k)^2},\nonumber\\
\mathcal{A_{\text{T2}}} &\propto \int_{k,l} \frac{\mathcal{F}_{T3}^{\alpha\beta}(l,k,p)}{k^2(k-p)^2 (k-l)^2 l^2(l-p)^2},\nonumber\\
\mathcal{A_{\text{T3}}} &\propto \int_{k,l} \frac{\mathcal{F}_{T4}^{\alpha\beta}(l,k,p)}{k^4(k-p)^2 l^2(l-k)^2},\nonumber\\
\mathcal{A_{\text{T4}}} &\propto \int_{k,l} \frac{\mathcal{F}_{T5}^{\alpha\beta}(l,k,p)}{k^2(k-p)^2 l^2(l-p)^2},\nonumber\\
\mathcal{A_{\text{T5}}} &\propto \int_{k,l} \frac{\mathcal{F}_{T6}^{\alpha\beta}(l,k,p)}{k^2 l^2 (l-k+p)^2}.
\end{align}   

\noindent
Notice that we are adopting Feynman gauge and scalars/fermions are massless as already pointed out. The equations above are already in normal form, meaning one can now apply the rules of IREG sketched in the last section. Another simplification of our case is that only the divergent part of the integrals above is needed for obtaining the gauge coupling $\beta$ function. In the next subsections we present the results of IREG for each diagram and for each of the theories we considered. 

In order to allow a comparison with dimensional regularisation methods, we perform Dirac and Lorentz algebra in $d$-dimensions and evaluate the integrals also in $d$-dimensions. 
We will keep track of terms coming from contractions of the metric $g_{\mu\nu}g^{\mu\nu}=d$, for instance, refraining to write $d=4-2\epsilon$ until the end of the calculation.
This will allow us to recover the results diagram-by-diagram obtained not only in CDR, but also in \textit{naive} DRED (devoid of $\epsilon$-scalars) which we will, hereafter, denote by $\overline{\text{DRED}}$. 
A similar approach was performed in \cite{EPSILONSCALAR-BETA}. In this way, we can define not only the minimal subtraction scheme in CDR ($\overline{\text{MS}}$) but also its analogue in $\overline{\text{DRED}}$ ($\overline{\text{DR}}$')\footnote{We will reserve the symbol $\overline{\text{DR}}$ to the  minimal subtaction scheme in the consistent version of DRED, with the  inclusion of $\epsilon$-scalars.}. Since both schemes are mass independent, the first two coefficients of the strong coupling $\beta$ function will be identical \cite{TARRACH}, a result we will recover. In section \ref{sec:espilon} we will include the $\epsilon$-scalars to define DRED properly. However, since we did not refrain to identify the multiplicity of these (fictitious) particles with $N_{\epsilon}=2\epsilon$, and only divergent terms are kept, we will recover the result of the $\overline{\text{DR}}$ scheme \cite{EPSILONSCALAR-BETA}, which agrees with our previous results in $\overline{\text{DR}}$', $\overline{\text{MS}}$. On the other hand, if we had performed a subtraction scheme removing terms $(N_{\epsilon}/\epsilon)^{n}$ as was done in \cite{BROGGIO}, the coefficients of the $\beta$ function would depend on $N_{\epsilon}$. Nevertheless, after identifying  $N_{\epsilon}=2\epsilon$, the same result will be recovered in the limit $\epsilon\rightarrow 0$.

Before proceeding to our explicit results, it is necessary to discuss briefly the renormalisation program in the case of the background field method. For definiteness, we will consider the Yang-Mills theory which has all the necessary ingredients. As discussed in \cite{Abbott}, in principle, one would need to renormalize the fields (gluon, background gluon, ghost), coupling, and, potentially, the gauge-fixing parameter. However, since only the background gluon field occurs in external legs, the renormalisation constants related to the gluon and ghost fields will always cancel. Therefore, one should only consider the multiplicative renormalisation constants defined below
\begin{equation}
\hat{A}_{o} = Z_{\hat{A}}^{1/2}\hat{A}_{r}; \quad g_{o} = Z_{g} g_{r}; \quad \alpha_{o} = Z_{\alpha} \alpha_{r},
\end{equation}
where $\hat{A}$, $g$, $\alpha$ stand for the background gluon field, strong coupling, and gauge-fixing parameter respectively. The latter could be left unrenormalized as well if we had considered a general gauge to perform the calculation. However, since we have adopted Feynman gauge throughout our work, it will be necessary to include counterterms related to gauge-fixing renormalisation. 

As standard, one can write
\begin{equation}
\alpha_{o} = Z_{\alpha} \alpha_{r} = (1+\delta_{\alpha})\alpha_{r},
\end{equation}
which implies
\begin{align}
\mathcal{L}_{\text{GF}}&=-\frac{1}{2\alpha_{o}}\left[\partial_{\mu}A^{\mu}_{a}+g_{o}f^{abc}(\hat{A}_{o})_{\mu}^{b}A_{\mu}^{c}\right]^{2}\nonumber\\
&=-\frac{1}{2\alpha_{r}}\left[\partial_{\mu}A^{\mu}_{a}+g_{r}f^{abc}(\hat{A}_{r})_{\mu}^{b}A_{\mu}^{c}\right]^{2}+\frac{\delta_{\alpha}}{2\alpha_{r}}\left[\partial_{\mu}A^{\mu}_{a}+g_{r}f^{abc}(\hat{A}_{r})_{\mu}^{b}A_{\mu}^{c}\right]^{2}
\end{align}
where the second term gives the counterterms related to the gauge-fixing renormalisation. Notice that we have used the important relation $Z_{g} = Z_{\hat{A}}^{-1/2}$ as well. Therefore, to obtain an explicit formula to $\delta_{\alpha}$ it suffices to compute the one-loop correction to the gluon propagator 
\begin{equation}
 \Pi_{\mu\nu}^{ab}|_{\text{div}}=-i5g^{2}C_{A}I_{log}(\lambda^{2})(g_{\mu\nu}p^{2}-p_{\mu}p_{\nu})\delta^{ab}\xrightarrow[\text{DRED}]{}\frac{5}{3\epsilon}\frac{g^{2}C_{A}}{(4\pi)^{2}}(g_{\mu\nu}p^{2}-p_{\mu}p_{\nu})\delta^{ab},
\end{equation}
since, given we are in Feynman gauge, we should attain the relation
\begin{equation}
\Pi_{\mu\nu}^{ab}|_{\text{div}}+\delta_{\alpha}p_{\mu}p_{\nu}\delta^{ab} \propto g_{\mu\nu}\delta^{ab}.   
\end{equation}
The final result is
\begin{equation}
\delta_{\alpha}=-i5g^{2}C_{A}I_{log}(\lambda^{2}) \xrightarrow[\text{DRED}]{}\frac{5}{3\epsilon}\frac{g^{2}C_{A}}{(4\pi)^{2}}
\end{equation}
which agrees with \cite{Abbott}.

% A very important relation that is enforced by the background field method is
% \begin{equation}
%     Z_{g} = Z_{\hat{A}}^{1/2},
% \end{equation}
% which mimics the relation among the photon field and gauge couling renormalisation constants presented in QED.

\subsection{Scalar QED}

In scalar QED, the part of the Lagrangian containing the couplings is of the form
\begin{equation}
\mathcal{L}_{\text{scalar QED}} \supset -ig_{e}A_{\mu}\left[\phi^{*}(\partial_{\mu}\phi)-(\partial_{\mu}\phi^{*})\phi\right] + g_{e}^{2}A_{\mu}A^{\mu}|\phi|^{2}.
\end{equation}

This implies that only topology T4 cannot be realised (remember the photon is in both external legs). Regarding counterterms, we only need to consider (possible) corrections due to renormalisation of the gauge fixing parameter. In the present case (abelian theory), none of the couplings will depend on this parameter, meaning we do not need to consider counterterms related to them. Moreover, since the triple coupling contains only one photon field, it is not possible to have a photon self-energy sub-diagram (see topology T1). Therefore, for our calculation of the $\beta$ function of scalar QED, no counterterm is needed. For completeness, we have indeed computed the counterterms related to the triple and quartic coupling (obtained from shrinking sub-diagrams in topologies T1 and T2 to a point), and the counterterm related to the scalar self-energy (obtained by shrinking the sub-diagram in topology T3 to a point). As expected, the counterterms cancel among themselves.

In order to present our results, we will define the two-loop correction to the photon field to be of the form
\begin{equation}
\label{eq: def A B}
\mathcal{A_{\mu\nu}}=\frac{i g_{e}^{4}}{(4\pi)^{4}} \left[A g_{\mu\nu}p^{2} - B p_{\mu}p_{\nu}\right].
\end{equation}
\noindent
where $p$ is the momentum carried by the external photon. With this definition at hand, the explicit results for each topology can be read from table \ref{tab:sca ireg} in the case of IREG and table \ref{tab:sca dreg} for CDR. In the case of scalar QED, there are no contractions of the type $g_{ab}g^{ab} = d$, which implies that results in CDR are identical to $\overline{\text{DRED}}$, as can also be seen from table \ref{tab:sca dreg}. We should also comment that in the case of scalar QED, there is only one type of diagram related  to each contributing topology, although topologies T1, T3 have multiplicity 4, 2 respectively. The results for the counterterms can be found in tables \ref{tab:ct sca ireg} and \ref{tab:ct sca dreg}, which cancel among themselves as already pointed out.

\begin{table}[h!]
\begin{equation}
\begin{array}{|c|c|c|c|c|c|c|c|c|}
\hline
\multicolumn{1}{ |c }{\multirow{2}{*}{\text{Topology}}}  & \multicolumn{4}{|c}{\text{A}} & \multicolumn{4}{|c|}{\text{B}} 
\\ \cline{2-9}
& I_{log}^{(2)}(\lambda^{2}) & I_{log}^{2}(\lambda^{2}) & \rho_{IREG} & I_{log}(\lambda^{2}) & I_{log}^{(2)}(\lambda^{2}) & I_{log}^{2}(\lambda^{2}) & \rho_{IREG} & I_{log}(\lambda^{2}) \\ \hline 
T1 & -\frac{2}{b} & \frac{2}{b^2} & -\frac{2}{b} & \frac{87}{9b} &  -\frac{2}{b} & \frac{2}{b^2} & -\frac{2}{b} & \frac{51}{9b}
\\ \hline 
T2 & \frac{2}{3b} & -\frac{2}{3b^2} & \frac{2}{3b} & -\frac{47}{9b} &  \frac{2}{3b} & -\frac{2}{3b^2} & \frac{2}{3b} & -\frac{35}{9b}
\\ \hline 
T3 & \frac{4}{3b} & -\frac{4}{3b^2} & \frac{4}{3b} & -\frac{58}{9b} & \frac{4}{3b} & -\frac{4}{3b^2} & \frac{4}{3b} & -\frac{52}{9b}
\\ \hline 
T5 & 0 & 0 & 0 & -\frac{2}{b} &  0 & 0 & 0 & 0
\\ \hline 
%\overline{T1} & 0 & \frac{4}{3b^2} & -\frac{4}{3b} & \frac{32}{9b} &  0 & \frac{4}{3b^2} & -\frac{4}{3b} & \frac{32}{9b}
%\\ \hline 
%\overline{T3} & 0 & -\frac{4}{3b^2} & \frac{4}{3b} & -\frac{32}{9b} &  0 & -\frac{4}{3b^2} & \frac{4}{3b} & -\frac{32}{9b}
%\\ \hline 
\text{Sum} & 0 & 0 & 0 & -\frac{4}{b} & 0 & 0 & 0 & -\frac{4}{b}
\\ \hline  
\end{array}\nonumber
\end{equation}
\caption{Results for Scalar QED using IREG where $\rho_{IREG} = I_{log}(\lambda^{2})\ln\left[-\frac{p^2}{\lambda^2}\right]$}
\label{tab:sca ireg}
\end{table}

\begin{table}
\begin{equation}
\begin{array}{|c|c|c|c|c|}
\hline
\text{Diagram}  & \text{A}_{\text{CDR}} & \text{A}_{\overline{\text{DRED}}}-\text{A}_{\text{CDR}} &\text{B}_{\text{CDR}} &  \text{B}_{\overline{\text{DRED}}}-\text{B}_{\text{CDR}} 
\\ \hline 
T1 & \frac{1}{\epsilon^{2}} + \frac{13-4\rho}{2\epsilon} & 0  & \frac{1}{\epsilon^{2}} + \frac{9-4\rho}{2\epsilon} & 0
\\ \hline
T2 & -\frac{1}{3\epsilon^{2}} - \frac{19-4\rho}{6\epsilon} & 0  & -\frac{1}{3\epsilon^{2}} - \frac{15-4\rho}{6\epsilon} & 0
\\ \hline 
T3 & -\frac{2}{3\epsilon^{2}} - \frac{13-4\rho}{3\epsilon} & 0 & -\frac{2}{3\epsilon^{2}} - \frac{12-4\rho}{3\epsilon} & 0
\\ \hline 
T5 & -\frac{1}{\epsilon} & 0  & 0 & 0
\\ \hline
%e & -\frac{4}{3\epsilon^{2}} - \frac{32-12\rho}{9\epsilon} & 0  & -\frac{4}{3\epsilon^{2}} - \frac{32-12\rho}{9\epsilon} & 0
%\\ \hline
%f & \frac{4}{3\epsilon^{2}} + \frac{32-12\rho}{9\epsilon} & 0  & \frac{4}{3\epsilon^{2}} + \frac{32-12\rho}{9\epsilon} & 0
%\\ \hline
\text{Sum} & -\frac{2}{\epsilon} & 0 & -\frac{2}{\epsilon} & 0
\\ \hline  
\end{array}\nonumber
\end{equation}
\caption{Results for Scalar QED using CDR and $\overline{\text{DRED}}$, where $\rho=\gamma_{E} - \ln 4\pi + \ln(p^2/\mu^{2}_{DR})$.}
\label{tab:sca dreg}
\end{table}

\begin{table}[h!]
\begin{equation}
\begin{array}{|c|c|c|c|c|c|c|c|c|}
\hline
\multicolumn{1}{ |c }{\multirow{2}{*}{\text{Counterterm}}}  & \multicolumn{4}{|c}{\text{A}} & \multicolumn{4}{|c|}{\text{B}} 
\\ \cline{2-9}
& I_{log}^{(2)}(\lambda^{2}) & I_{log}^{2}(\lambda^{2}) & \rho_{IREG} & I_{log}(\lambda^{2}) & I_{log}^{(2)}(\lambda^{2}) & I_{log}^{2}(\lambda^{2}) & \rho_{IREG} & I_{log}(\lambda^{2}) \\ \hline 
\text{Coupling} & 0 & \frac{4}{3b^2} & -\frac{4}{3b} & \frac{32}{9b} &  0 & \frac{4}{3b^2} & -\frac{4}{3b} & \frac{32}{9b}
\\ \hline 
\text{Scalar self-energy} & 0 & -\frac{4}{3b^2} & \frac{4}{3b} & -\frac{32}{9b} &  0 & -\frac{4}{3b^2} & \frac{4}{3b} & -\frac{32}{9b}
\\ \hline 
\text{Sum} & 0 & 0 & 0 & 0 & 0 & 0 & 0 & 0
\\ \hline  
\end{array}\nonumber
\end{equation}
\caption{Counterterm results for Scalar QED using IREG where $\rho_{IREG} = I_{log}(\lambda^{2})\ln\left[-\frac{p^2}{\lambda^2}\right]$}
\label{tab:ct sca ireg}
\end{table}

\begin{table}
\begin{equation}
\begin{array}{|c|c|c|c|c|}
\hline
\text{Counterterm}  & \text{A}_{\text{CDR}} & \text{A}_{\overline{\text{DRED}}}-\text{A}_{\text{CDR}} &\text{B}_{\text{CDR}} &  \text{B}_{\overline{\text{DRED}}}-\text{B}_{\text{CDR}} 
\\ \hline 
\text{Coupling} & \frac{4}{3\epsilon^{2}} + \frac{32-12\rho}{9\epsilon} & 0  & \frac{4}{3\epsilon^{2}} + \frac{32-12\rho}{9\epsilon} & 0
\\ \hline
\text{Scalar self-energy} & -\frac{4}{3\epsilon^{2}} - \frac{32-12\rho}{9\epsilon} & 0  & -\frac{4}{3\epsilon^{2}} - \frac{32-12\rho}{9\epsilon} & 0
\\ \hline

\text{Sum} & 0 & 0 & 0 & 0
\\ \hline  
\end{array}\nonumber
\end{equation}
\caption{Counterterm results for Scalar QED using CDR and $\overline{\text{DRED}}$, where $\rho=\gamma_{E} - \ln 4\pi + \ln(p^2/\mu^{2}_{DR})$}
\label{tab:ct sca dreg}
\end{table}

\newpage
As one can easily notice, the end result is gauge invariant in all of the methods considered here. Also, as is well-known \cite{Schwartz}, although we are considering a two-loop correction, only terms proportional to $\epsilon^{-1}$ will survive in the end result. In the framework of IREG we also have a similar pattern, since only $I_{log}(\lambda^{2})$ terms appear in the final result. This similarity between methods, however, will not be valid in general, as we are going to see when studying the Yang-Mills theory. The reason can be traced back to the appearance of diagrams of topology T4, as we will explain later.

\subsection{Spinorial QED}

We move to spinorial QED, where, although only one type of vertex occurs $-e\bar{\psi}\gamma^{\mu}\psi A_{\mu}$, one needs to deal with Dirac algebra. Given the more restrictive coupling, when compared with scalar QED, one expects that even less diagrams will contribute.  Explicitly, only topologies T2 and T3 can be realized, each one with one diagram, and multiplicity 1,2 respectively. As before, no counterterms will be necessary. For completeness, we compute them and check that they cancel. 

Regarding the computation itself, the only point that one should be careful about is in how the Dirac algebra is performed. Now, spurious terms $\mathcal{O}(\epsilon)$ will appear, implying that the results in CDR and $\overline{\text{DRED}}$ are not identical diagram by diagram, although the final result should be. We checked that this is indeed the case. In the framework of IREG, some care should also be exercised, in order to perform tensor identification consistently. As we argued in section \ref{sec:rules}, a normal form is obtained after performing Dirac and Lorentz algebra to the whole diagram. After the normal is attained, manipulations in the numerator that may generate spurious terms with $k^2$, for instance, cannot be performed. An example of such \textit{forbidden} manipulation is 
\begin{align}
\int_{k,q}\frac{k.q}{k^2 (k-p)^2 q^2 (q-p)^2 (k-q)^2}\bigg|_{\text{IREG}}\neq \int_{k,q}\frac{k^2 + q^2 - (k-q)^2}{2 k^2 (k-p)^2 q^2 (q-p)^2 (k-q)^2}\bigg|_{\text{IREG}}.
\end{align}
The left-hand side of the equation above appears in the diagram of topology T2, for instance. The consistent approach to treat this integral is given by the procedure defined in \cite{ADRIANO}, whose result we collect in the appendix \ref{app: integrals}. 

To conclude this subsection we collect our results in tables \ref{tab:qed ireg} to \ref{tab: ct qed dreg}, adopting the same convention of eq. \ref{eq: def A B}. As in the case of scalar QED, the end result is transverse in all methods and depends only in $\epsilon^{-1}$ or $I_{log}(\lambda^{2})$ terms. Moreover, the counterterms cancel among themselves, and we notice that the two-loop correction to the photon is the same in both scalar and spinorial QED. This implies the same gauge coupling $\beta$ function.

\begin{table}[h!]
\begin{equation}
\begin{array}{|c|c|c|c|c|c|c|c|c|}
\hline
\multicolumn{1}{ |c }{\multirow{2}{*}{\text{Topology}}}  & \multicolumn{4}{|c}{\text{A}} & \multicolumn{4}{|c|}{\text{B}} 
\\ \cline{2-9}
& I_{log}^{(2)}(\lambda^{2}) & I_{log}^{2}(\lambda^{2}) & \rho_{IREG} & I_{log}(\lambda^{2}) & I_{log}^{(2)}(\lambda^{2}) & I_{log}^{2}(\lambda^{2}) & \rho_{IREG} & I_{log}(\lambda^{2}) \\ \hline 
T2 & \frac{8}{3b} & -\frac{8}{3b^2} & \frac{8}{3b} & \frac{92}{9b} &  \frac{8}{3b} & -\frac{8}{3b^2} & \frac{8}{3b} & \frac{104}{9b}
\\ \hline
T3 & -\frac{8}{3b} & \frac{8}{3b^2} & -\frac{8}{3b} & -\frac{128}{9b} & -\frac{8}{3b} & \frac{8}{3b^2} & -\frac{8}{3b} & -\frac{140}{9b}
\\ \hline 
\text{Sum} & 0 & 0 & 0 & -\frac{4}{b} & 0 & 0 & 0 & -\frac{4}{b}
\\ \hline  
\end{array}\nonumber
\end{equation}
\caption{Results for Spinorial QED using IREG where $\rho_{IREG} = I_{log}(\lambda^{2})\ln\left[-\frac{p^2}{\lambda^2}\right]$}
\label{tab:qed ireg}
\end{table}

\begin{table}[h!]
\begin{equation}
\begin{array}{|c|c|c|c|c|}
\hline
\text{Diagram}  & \text{A}_{\text{CDR}} & \text{A}_{\overline{\text{DRED}}}-\text{A}_{\text{CDR}} &\text{B}_{\text{CDR}} &  \text{B}_{\overline{\text{DRED}}}-\text{B}_{\text{CDR}} 
\\ \hline 
T2 & \frac{4}{3\epsilon^{2}} + \frac{8(2-\rho)}{3\epsilon} & \frac{4}{3\epsilon}  & \frac{4}{3\epsilon^{2}} + \frac{2(7-4\rho)}{3\epsilon} & \frac{4}{3\epsilon}
\\ \hline
T3 & -\frac{4}{3\epsilon^{2}} - \frac{2(11-4\rho)}{3\epsilon} & -\frac{4}{3\epsilon} & -\frac{4}{3\epsilon^{2}} - \frac{4(5-2\rho)}{3\epsilon} & -\frac{4}{3\epsilon}
\\ \hline 
\text{Sum} & -\frac{2}{\epsilon} & 0 & -\frac{2}{\epsilon} & 0
\\ \hline  
\end{array}\nonumber
\end{equation}
\caption{Results for Spinorial QED using CDR and $\overline{\text{DRED}}$, where $\rho=\gamma_{E} - \ln 4\pi + \ln(p^2/\mu^{2}_{DR})$}
\label{tab: qed dreg}
\end{table}

\begin{table}[h!]
\begin{equation}
\begin{array}{|c|c|c|c|c|c|c|c|c|}
\hline
\multicolumn{1}{ |c }{\multirow{2}{*}{\text{Counterterm}}}  & \multicolumn{4}{|c}{\text{A}} & \multicolumn{4}{|c|}{\text{B}} 
\\ \cline{2-9}
& I_{log}^{(2)}(\lambda^{2}) & I_{log}^{2}(\lambda^{2}) & \rho_{IREG} & I_{log}(\lambda^{2}) & I_{log}^{(2)}(\lambda^{2}) & I_{log}^{2}(\lambda^{2}) & \rho_{IREG} & I_{log}(\lambda^{2}) \\ \hline 
\text{Coupling} & 0 & \frac{8}{3b^2} & -\frac{8}{3b} & \frac{40}{9b} &  0 & \frac{8}{3b^2} & -\frac{8}{3b} & \frac{40}{9b}
\\ \hline  
\text{Scalar self-energy} & 0 & -\frac{8}{3b^2} & \frac{8}{3b} & -\frac{40}{9b} &  0 & -\frac{8}{3b^2} & \frac{8}{3b} & -\frac{40}{9b}
\\ \hline
\text{Sum} & 0 & 0 & 0 & 0 & 0 & 0 & 0 & 0
\\ \hline  
\end{array}\nonumber
\end{equation}
\caption{Counterterm results for Spinorial QED using IREG where $\rho_{IREG} = I_{log}(\lambda^{2})\ln\left[-\frac{p^2}{\lambda^2}\right]$}
\label{tab:ct qed ireg}
\end{table}

\begin{table}
\begin{equation}
\begin{array}{|c|c|c|c|c|}
\hline
\text{Counterterm}  & \text{A}_{\text{CDR}} & \text{A}_{\overline{\text{DRED}}}-\text{A}_{\text{CDR}} &\text{B}_{\text{CDR}} &  \text{B}_{\overline{\text{DRED}}}-\text{B}_{\text{CDR}} 
\\ \hline 
\text{Coupling} & \frac{8}{3\epsilon^{2}} + \frac{8(7-3\rho)}{9\epsilon} & 0  & \frac{8}{3\epsilon^{2}} + \frac{8(7-3\rho)}{9\epsilon} & 0
\\ \hline 
\text{Scalar self-energy} & -\frac{8}{3\epsilon^{2}} - \frac{8(7-3\rho)}{9\epsilon} & 0  & -\frac{8}{3\epsilon^{2}} - \frac{8(7-3\rho)}{9\epsilon} & 0
\\ \hline 
\text{Sum} & 0 & 0 & 0 & 0
\\ \hline  
\end{array}\nonumber
\end{equation}
\caption{Counterterm results for Spinorial QED using CDR and $\overline{\text{DRED}}$, where $\rho=\gamma_{E} - \ln 4\pi + \ln(p^2/\mu^{2}_{DR})$}
\label{tab: ct qed dreg}
\end{table}

\newpage

\subsection{Pure Yang-Mills}

\begin{figure}
\centering
\subcaptionbox{}{\includegraphics[width=0.20\textwidth]{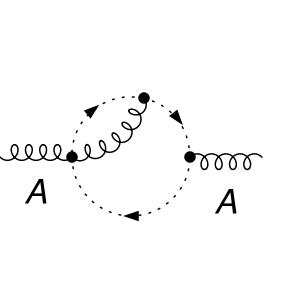}}%
\hfill
\subcaptionbox{}{\includegraphics[width=0.20\textwidth]{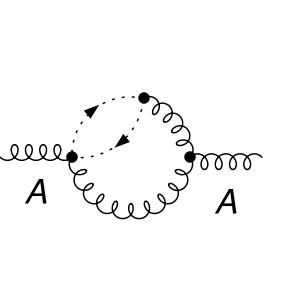}}%
\hfill
\subcaptionbox{}{\includegraphics[width=0.20\textwidth]{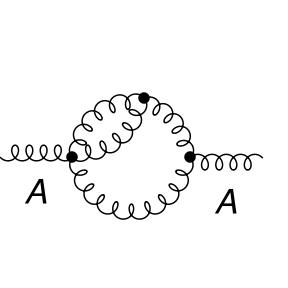}}%
\hfill
\subcaptionbox{}{\includegraphics[width=0.20\textwidth]{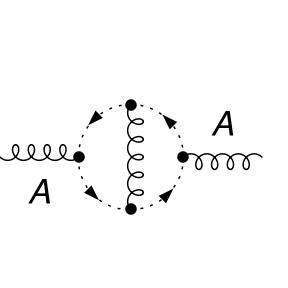}}%
\vfill
\subcaptionbox{}{\includegraphics[width=0.20\textwidth]{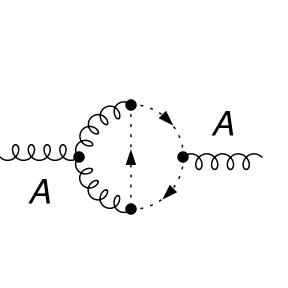}}%
\hfill
\subcaptionbox{}{\includegraphics[width=0.20\textwidth]{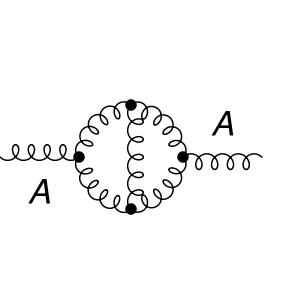}}%
\hfill
\subcaptionbox{}{\includegraphics[width=0.20\textwidth]{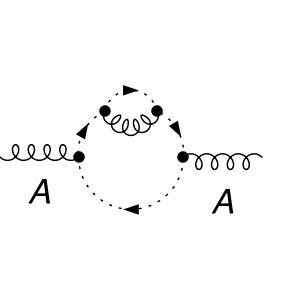}}%
\hfill
\subcaptionbox{}{\includegraphics[width=0.20\textwidth]{fig16}}%
\vfill
\subcaptionbox{}{\includegraphics[width=0.20\textwidth]{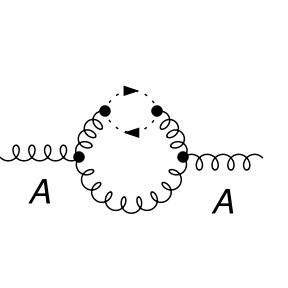}}%
\hfill
\subcaptionbox{}{\includegraphics[width=0.20\textwidth]{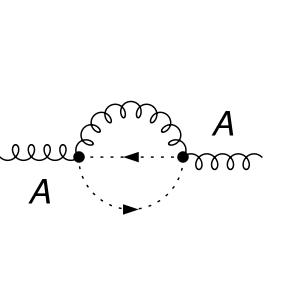}}%
\hfill
\subcaptionbox{}{\includegraphics[width=0.20\textwidth]{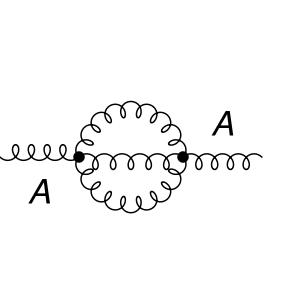}}%
\hfill
\subcaptionbox{}{\includegraphics[width=0.20\textwidth]{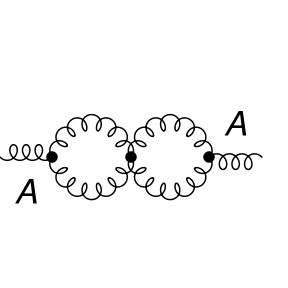}}%
\caption{Two-loop correction to the two-point function of the background field $A$.}
\label{Feynman}
\end{figure}

We turn to non-abelian theories. At first, we do not include scalar/fermionic interactions, considering pure Yang-Mills theory, see Appendix \ref{app:feynman} for notations and conventions of the Feynman rules used in the background field method to Yang-Mills \cite{Abbott}, employed in the present calculation. The calculation proves to be involved enough since not only all topologies are realized, but also there are diagrams with different field content for some of the topologies, as can be seen in fig. \ref{Feynman}.

Apart from the large number of diagrams, there are two main differences regarding the computation in QED we would like to emphasize. First, we have the appearance of topology T5 which, as we are going to see, implies in the presence of not only terms with $I_{log}(\lambda^{2})$ in the end result, but also $I_{log}^{(2)}(\lambda^{2})$, and $I_{log}^{2}(\lambda^{2})$. Second, not only the triple coupling depends on the gauge fixing parameter, but also the 1-loop correction to the gluon self-energy appears as sub-diagram in diagrams (h) and (i). Therefore, we will need to consider counterterms related to the gauge fixing renormalisation. 

In a similar manner to eq. \ref{eq: def A B}, we can define
\begin{equation}
\frac{i g_{s}^{4} C_{A}^{2} \delta^{ab}}{(4\pi)^{4}} \left[A g_{\mu\nu}p^{2} - B p_{\mu}p_{\nu}\right],
\label{eq: def A B gs}
\end{equation}
\noindent
where $p$ is the external momenta carried by the background field. Regarding the diagrams of Fig. \ref{Feynman}, our results, for IREG, are presented in table \ref{tab: YM ireg} while for CDR/$\overline{\text{DRED}}$ they can be seen in table \ref{tab: YM dreg}. 

\begin{table}
\begin{equation}
\begin{array}{|c|c|c|c|c|c|c|c|c|}
\hline
\multicolumn{1}{ |c }{\multirow{2}{*}{\text{Diagram}}}  & \multicolumn{4}{|c}{\text{A}} & \multicolumn{4}{|c|}{\text{B}} 
\\ \cline{2-9}
& I_{log}^{(2)}(\lambda^{2}) & I_{log}^{2}(\lambda^{2}) & \rho_{IREG} & I_{log}(\lambda^{2}) & I_{log}^{(2)}(\lambda^{2}) & I_{log}^{2}(\lambda^{2}) & \rho_{IREG} & I_{log}(\lambda^{2}) \\ \hline 
a & \frac{1}{3b} & -\frac{1}{3b^2} & \frac{1}{3b} & -\frac{29}{18b} & \frac{1}{3b} & -\frac{1}{3b^2} & \frac{1}{3b} & -\frac{17}{18b}
\\ \hline 
b & \frac{5}{12b} & -\frac{5}{12b^2} & \frac{5}{12b} & -\frac{97}{72b} & \frac{5}{12b} & -\frac{5}{12b^2} & \frac{5}{12b} & -\frac{109}{72b}
\\ \hline 
c & \frac{9}{4b} & -\frac{9}{4b^2} & \frac{9}{4b} & -\frac{39}{8b} & \frac{9}{4b} & -\frac{9}{4b^2} & \frac{9}{4b} & -\frac{75}{8b}
\\ \hline 
d & -\frac{1}{12b} & \frac{1}{12b^2} & -\frac{1}{12b} & \frac{47}{72b}& -\frac{1}{12b} & \frac{1}{12b^2} & -\frac{1}{12b} & \frac{35}{72b}
\\ \hline 
e & \frac{1}{2b} & -\frac{1}{2b^2} & \frac{1}{2b} & -\frac{7}{4b} &\frac{1}{2b} & -\frac{1}{2b^2} & \frac{1}{2b} & -\frac{7}{4b} 
\\ \hline 
f & -\frac{27}{4b} & \frac{27}{4b^2} & -\frac{27}{4b} & \frac{195}{8b} & -\frac{27}{4b} & \frac{27}{4b^2} & -\frac{27}{4b} & \frac{207}{8b}
\\ \hline 
g & -\frac{1}{3b} & \frac{1}{3b^2} & -\frac{1}{3b} & \frac{29}{18b} & -\frac{1}{3b} & \frac{1}{3b^2} & -\frac{1}{3b} & \frac{13}{9b}
\\ \hline 
h+i & -\frac{25}{3b} & \frac{25}{3b^2} & -\frac{25}{3b} & \frac{521}{18b} &  -\frac{25}{3b} & \frac{25}{3b^2} & -\frac{25}{3b} & \frac{268}{9b}
\\ \hline 
j & 0 & 0 & 0 & \frac{1}{4b} & 0 & 0 & 0  & 0
\\ \hline 
k & 0 & 0 & 0 & -\frac{9}{4b}& 0 & 0 & 0 & 0
\\ \hline 
l & 0 & -\frac{6}{b^2} & \frac{12}{b} & -\frac{24}{b} &  0 & -\frac{6}{b^2} & \frac{12}{b} & -\frac{24}{b}
\\ \hline 
\text{Sum} & -\frac{12}{b} & \frac{6}{b^2} & 0 & \frac{20}{b} & -\frac{12}{b} & \frac{6}{b^2} & 0 & \frac{20}{b}
\\ \hline  
\end{array}\nonumber
\end{equation}
\caption{Results for pure Yang-Mills using IREG where $\rho_{IREG} = I_{log}(\lambda^{2})\ln\left[-\frac{p^2}{\lambda^2}\right]$}
\label{tab: YM ireg}
\end{table}

\begin{table}
\begin{equation}
\begin{array}{|c|c|c|c|c|}
\hline
\text{Diagram}  & \text{A}_{\text{CDR}} & \text{A}_{\overline{\text{DRED}}}-\text{A}_{\text{CDR}} &\text{B}_{\text{CDR}} &  \text{B}_{\overline{\text{DRED}}}-\text{B}_{\text{CDR}} 
\\ \hline 
a & -\frac{1}{6\epsilon^{2}} + \frac{-13+4\rho}{12\epsilon} & 0 & -\frac{1}{6\epsilon^{2}} + \frac{-9+4\rho}{12\epsilon} & 0
\\ \hline 
b & -\frac{5}{24\epsilon^{2}} + \frac{-41+20\rho}{48\epsilon} & 0 & -\frac{5}{24\epsilon^{2}} + \frac{5(-9+4\rho)}{48\epsilon} & 0
\\ \hline 
c & -\frac{9}{8\epsilon^{2}} + \frac{-57+36\rho}{16\epsilon} & \frac{3}{4\epsilon} & -\frac{9}{8\epsilon^{2}} + \frac{-93+36\rho}{16\epsilon} & \frac{3}{4\epsilon}
\\ \hline 
d & \frac{1}{24\epsilon^{2}} + \frac{19-4\rho}{48\epsilon} & 0 & \frac{1}{24\epsilon^{2}} + \frac{15-4\rho}{48\epsilon} & 0
\\ \hline 
e & -\frac{1}{4\epsilon^{2}} + \frac{-9+4\rho}{8\epsilon} & 0 & -\frac{1}{4\epsilon^{2}} + \frac{-9+4\rho}{8\epsilon} & 0
\\ \hline 
f & \frac{27}{8\epsilon^{2}} + \frac{233-108\rho}{16\epsilon} & \frac{3}{4\epsilon} & \frac{27}{8\epsilon^{2}} + \frac{245-108\rho}{16\epsilon} & \frac{3}{4\epsilon}
\\ \hline 
g & \frac{1}{6\epsilon^{2}} + \frac{13-4\rho}{12\epsilon} & 0 & \frac{1}{6\epsilon^{2}} + \frac{3-\rho}{3\epsilon} & 0
\\ \hline 
h+i & \frac{25}{6\epsilon^{2}} + \frac{215-100\rho}{12\epsilon} & -\frac{3}{2\epsilon}  & \frac{25}{6\epsilon^{2}} + \frac{110-50\rho}{6\epsilon} & -\frac{3}{2\epsilon}
\\ \hline 
j & \frac{1}{8\epsilon} & 0 & 0 & 0
\\ \hline 
k & -\frac{9}{8\epsilon} & 0 & 0 & 0
\\ \hline 
l & -\frac{6}{\epsilon^{2}} + \frac{12(-2+\rho)}{\epsilon} & 0 & -\frac{6}{\epsilon^{2}} + \frac{12(-2+\rho)}{\epsilon} & 0
\\ \hline 
\text{Sum} & \frac{7}{3\epsilon} & 0 & \frac{7}{3\epsilon} & 0
\\ \hline  
\end{array}\nonumber
\end{equation}
\caption{Results for pure Yang-Mills using CDR and $\overline{\text{DRED}}$, where $\rho=\gamma_{E} - \ln 4\pi + \ln(p^2/\mu^{2}_{DR})$}
\label{tab: YM dreg}
\end{table}

\newpage

Some comments are in order. First notice in the IREG results that, apart from minus signs and/or global factors encoded in the parameter $b$, diagrams (a) to (i) have the same coefficients for $I_{log}^{(2)}(\lambda^{2})$, $I_{log}^{2}(\lambda^{2})$, and $\rho_{IREG}= I_{log}(\lambda^{2})\ln(-p^2/\lambda^2)$. This fact can be schematically understood as described below. Consider that the amplitude of one of those diagrams is given by 
\begin{align}
\mathcal{A} = \int_{k,l} \mathcal{F}(l,k,p). 
\end{align}
\noindent
Guided by the procedure presented in \cite{ADRIANO}, it is possible to separate the $F(l,k,p)$ function in different pieces, organizing the order of the integration in $l,k$ for each of them. Suppose that one of these terms is given as below, where one must first perform the integration in $l$, then in $k$
\begin{align}
\mathcal{A} \supset&\int_{k} G(k,p,\mu^{2}) \int_{l} F(l,k,p,\mu^{2}).
\end{align}
We have also included the $\mu^{2}$ parameter in denominators as explained in section \ref{sec:rules}. At this point, one should identify the BDI presented in the $l$ integral, encoded as $I_{log}(\mu^{2})$, and apply the scale relation to trade $\mu^{2}$ by $\lambda^{2}$
\begin{align}
\mathcal{A} \supset& \int_{k} G(k,p,\mu^{2}) \left[ a_{1} I_{log}(\mu^{2}) + a_{2}\right],\nonumber\\
=& \int_{k} G(k,p,\mu^{2}) \left[ a_{1} I_{log}(\lambda^{2}) - a_{1} b\ln\left[-\frac{k^2}{\lambda^2}\right]+
 \bar{a}_{2}\right],
\end{align}
where the actual form of $a_{2}$, $\bar{a}_{2}$ are not relevant here.
At this point one has to proceed to the integral in $k$. Adopting a similar treatment as done in the $l$ integral one obtains  
\begin{align}
\mathcal{A} \supset\,& a_{1} I_{log}(\lambda^{2})\int_{k} G(k,p,\mu^{2})-a_{1}b\int_{k} G(k,p,\mu^{2})\ln\left[-\frac{k^2}{\lambda^2}\right]\nonumber\\
=& a_{1} I_{log}(\lambda^{2})\left[A_{1}I_{log}(\mu^{2}) + \cdots \right]-a_{1}b\left[A_{1}I_{log}^{(2)}(\mu^{2}) + \cdots\right]\nonumber\\
=& a_{1}A_{1}\left[I_{log}^{2}(\lambda^{2})- bI_{log}(\lambda^{2})\ln\left[-\frac{p^2}{\lambda^2}\right] -bI_{log}^{(2)}(\lambda^{2}) + \cdots\right]
\label{eq: pattern}
\end{align}
where we made use of the relation
\begin{align}
\int_{k} G(k,p,\mu^{2}) \ln^{n}\left[-\frac{k^2}{\lambda^2}\right] = A_{1} I_{log}^{(n+1)}(\mu^{2}) + \cdots
\end{align}
whose proof we perform in Appendix \ref{app: proof}.

The important lesson to be taken from eq. \ref{eq: pattern} is that the coefficients of $I_{log}^{(2)}(\lambda^{2})$, $I_{log}^{2}(\lambda^{2})$, and $\rho_{IREG}$ are correlated. Explicitly, there is a minus sign and $b$ factor difference among $I_{log}^{(2)}(\lambda^{2})$, $\rho_{IREG}$, and $I_{log}^{2}(\lambda^{2})$, which reproduces the pattern we found for diagrams a to i, see table \ref{tab: YM ireg}. Notice that there is no such correlation for the $I_{log}(\lambda^{2})$ terms. 

The above reasoning cannot be applied when dealing with diagrams of topology T5, since, in this case, the integrals in $l,k$ are independent. Schematically,
\begin{align}
\mathcal{B} =& \int_{k} G(k,p)\int_{l} F(l,p)\nonumber\\
=& \left[B_{1}I_{log}(\mu^{2}) + \cdots \right]\left[C_{1}I_{log}(\mu^{2}) + \cdots \right]\nonumber\\
=&\left[B_{1}I_{log}(\lambda^{2})- B_{1} b\ln\left[-\frac{p^2}{\lambda^2}\right]+\cdots\right]\left[C_{1}I_{log}(\lambda^{2})- C_{1} b\ln\left[-\frac{p^2}{\lambda^2}\right]+\cdots\right]\nonumber\\
=& B_{1}C_{1}\left[I_{log}^{2}(\lambda^{2})- 2b\rho_{IREG}\right] + \cdots
\label{eq: pattern ilog2}
\end{align}
\noindent
a pattern which can once again be read from the diagram l of table \ref{tab: YM ireg}. Therefore, it is clear that the appearance of topology T5 will (potentially) break the pattern found before among $I_{log}^{(2)}(\lambda^{2})$, $I_{log}^{2}(\lambda^{2})$, and $\rho_{IREG}$ in the end result. Actually, this can be seen also in table \ref{tab: YM ireg}, where the sum of the results is void of $\rho_{IREG}$, while both $I_{log}^{(2)}(\lambda^{2})$, $I_{log}^{2}(\lambda^{2})$ are still present. This fact also explains why in the QED case only terms proportional to $I_{log}(\lambda^{2})$ survive, since it is not possible to realize topology T5 there.

Regarding dimensional methods, this distinction is not present. In other words, there is a correlation among $\epsilon^{-2}$ and the $\rho$ coefficient for all topologies. Therefore, since the end result is local (void of $\rho$ terms as in the IREG case), no term proportional to $\epsilon^{-2}$ can survive, as can be seen in table \ref{tab: YM dreg}. It can also be noticed that there are some differences among CDR and $\overline{\text{DRED}}$ in diagrams (c), (f), (h), although their sum vanishes. The reason can be traced back to contractions of the form $g_{ab}g^{ab} = d$, which may generate terms of order $\epsilon^{-1}$ when present in two-loop diagrams. This feature was already observed in spinorial QED. As explained in section \ref{sec:rules}, a consistent treatment of DRED requires the introduction of contributions with $\epsilon$-scalars, which, in view of the above results, must conspire to cancel among themselves. We will show this explicitly in the next subsection.

Finally, we notice that our end result is already gauge invariant\footnote{Actually, the terms proportional to $\rho_{IREG}$; $\rho$ (which are correlated to $I_{log}^{2}(\lambda^{2})$,$I_{log}^{(2)}(\lambda^{2})$; $\epsilon^{-2}$) are gauge invariant diagram by diagram, in accordance to the findings of \cite{Abbott}.} and local (non-local terms are encoded in $\rho$ or $\rho_{IREG}$, which vanish as already pointed out), although
we didn't include any counterterms yet. This implies that any gauge breaking or non-local terms appearing in the counterterms must cancel among themselves. Also, with a similar reasoning regarding eq. \ref{eq: pattern ilog2}, it is not hard to convince oneself that, apart from $\rho_{IREG}$, only terms with $I_{log}^{2}(\lambda^{2})$ or $I_{log}(\lambda^{2})$ can appear in the counterterms and the end result must be independent of  $I_{log}^{2}(\lambda^{2})$. This is indeed the case as shown in table \ref{tab: ct YM ireg} for IREG and table \ref{tab: ct YM dreg} for dimensional methods. We should emphasize that our results for CDR exactly reproduce the ones found in previous works \cite{Abbott}.

\begin{table}
\begin{equation}
\begin{array}{|c|c|c|c|c|c|c|c|c|}
\hline
\multicolumn{1}{ |c }{\multirow{2}{*}{\text{Counterterm}}}  & \multicolumn{4}{|c}{\text{A}} & \multicolumn{4}{|c|}{\text{B}} 
\\ \cline{2-9}
& I_{log}^{(2)}(\lambda^{2}) & I_{log}^{2}(\lambda^{2}) & \rho_{IREG} & I_{log}(\lambda^{2}) & I_{log}^{(2)}(\lambda^{2}) & I_{log}^{2}(\lambda^{2}) & \rho_{IREG} & I_{log}(\lambda^{2}) \\ \hline 
AA\hat{A} \text{ Coupling} & 0 & -\frac{25}{9b} & \frac{25}{9b} & -\frac{140}{27b} &  0 & -\frac{25}{9b} & \frac{25}{9b} & -\frac{140}{27b}
\\ \hline  
\text{Gluon self-energy} & 0 & \frac{25}{9b} & -\frac{25}{9b}  & \frac{230}{27b} & 0 & \frac{25}{9b} & -\frac{25}{9b}  & \frac{230}{27b}
\\ \hline 
\text{Sum} & 0 & 0 & 0 & \frac{10}{3b} & 0 & 0 & 0 & \frac{10}{3b}
\\ \hline  
\end{array}\nonumber
\end{equation}
\caption{Counterterms results for pure Yang-Mills using IREG where $\rho_{IREG} = I_{log}(\lambda^{2})\ln\left[-\frac{p^2}{\lambda^2}\right]$}
\label{tab: ct YM ireg}
\end{table}

\begin{table}
\begin{equation}
\begin{array}{|c|c|c|c|c|}
\hline
\text{Counterterm}  & \text{A}_{\text{CDR}} & \text{A}_{\text{DRED}}-\text{A}_{\text{CDR}} &\text{B}_{\text{CDR}} &  \text{B}_{\text{DRED}}-\text{B}_{\text{CDR}} 
\\ \hline 
AA\hat{A} \text{ Coupling} & -\frac{25}{9\epsilon^{2}} + \frac{5(-28+15\rho)}{27\epsilon} & 0 & -\frac{25}{9\epsilon^{2}} + \frac{5(-28+15\rho)}{27\epsilon} & 0
\\ \hline  
\text{Gluon self-energy} & \frac{25}{9\epsilon^{2}} + \frac{5(46-15\rho)}{27\epsilon} & 0 & \frac{25}{9\epsilon^{2}} + \frac{5(46-15\rho)}{27\epsilon} & 0
\\ \hline 
\text{Sum} & \frac{10}{3\epsilon} & 0 & \frac{10}{3\epsilon} & 0
\\ \hline  
\end{array}\nonumber
\end{equation}
\caption{Counterterms results for pure Yang-Mills using CDR and $\overline{\text{DRED}}$, where $\rho=\gamma_{E} - \ln 4\pi + \ln(p^2/\mu^{2}_{DR})$}
\label{tab: ct YM dreg}
\end{table}

%\newpage

\subsection{$\epsilon$-scalars for the YM theory}
\label{sec:espilon}

In the previous subsection we showed that some of the diagrams have different divergent parts when comparing a naive DRED scheme (without $\epsilon$-scalars) and CDR. However, the sum of the contributions is the same, regardless of the scheme chosen as expected (the two-loop coefficient of the beta function is the same in mass-independent renormalisation schemes). This observation implicitly shows that the contributions of $\epsilon$-scalars must conspire to render the same divergent part in the end in both schemes. In this subsection we explicitly show that this is the case.

\begin{figure}
\centering
\subcaptionbox*{c1}{\includegraphics[width=0.20\textwidth]{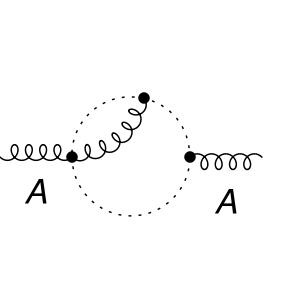}}%
\hfill
\subcaptionbox*{c2}{\includegraphics[width=0.20\textwidth]{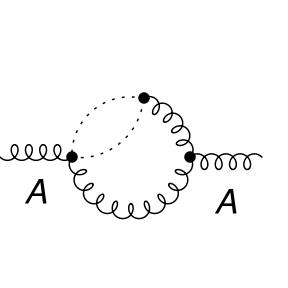}}%
\hfill
\subcaptionbox*{f1}{\includegraphics[width=0.20\textwidth]{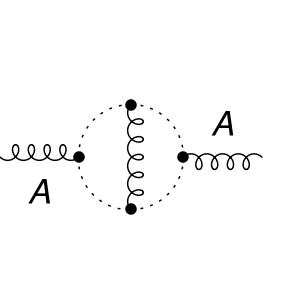}}%
\hfill
\subcaptionbox*{f2}{\includegraphics[width=0.20\textwidth]{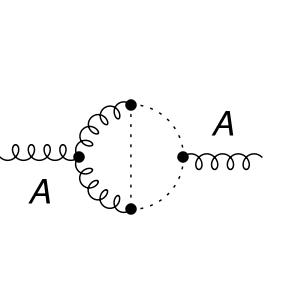}}%
\vfill
\subcaptionbox*{h1}{\includegraphics[width=0.20\textwidth]{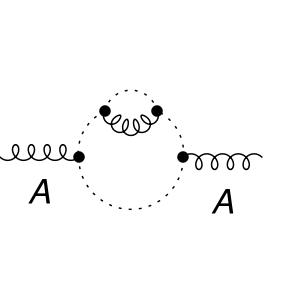}}%
\hfill
\subcaptionbox*{h2}{\includegraphics[width=0.20\textwidth]{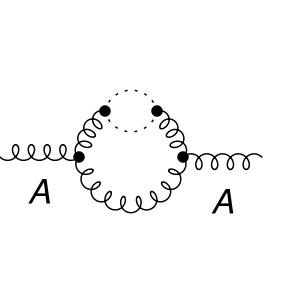}}%
\hfill
\subcaptionbox*{k}{\includegraphics[width=0.20\textwidth]{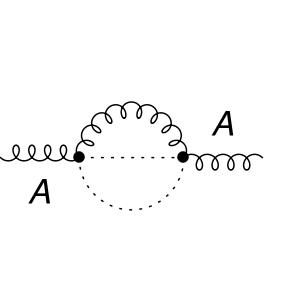}}%
\hfill
\subcaptionbox*{l}{\includegraphics[width=0.20\textwidth]{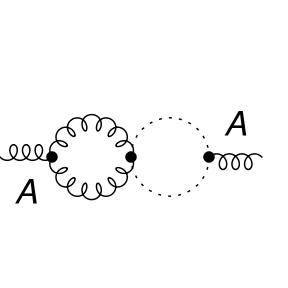}}%
\caption{Two-loop corrections with $\epsilon$-scalar. }
\label{fig:eps}
\end{figure}

As is well-known, $\epsilon$-scalars must be included in DRED for consistency \cite{EPSILONSCALAR}. They occur as a split of the gauge field as
\begin{equation}
A_{\mu}|_{4} = A_{\mu}|_{4-2\epsilon} + A_{\mu}|_{2\epsilon}
\end{equation} 
where the last term is the $\epsilon$-scalar.
Therefore, they will certainly occur in diagrams that contain only gluons, namely (c), (f), (h), (k), (l) from Fig. \ref{Feynman}. The diagrams containing $\epsilon$-scalars are depicted in Fig. \ref{fig:eps}. The contributions  related to diagram (l) vanish while the one related to diagram (k) will be of order $\rm{O}(\epsilon^{0})$, not relevant for our purposes. Therefore, one only needs the contribution from the other three types of diagrams. 

Adopting the same notation of eq. \ref{eq: def A B gs}, the results are collected in table \ref{tab:epsilon}. They are all gauge invariant, and cancel as they should. More interestingly, adding the results of the $\epsilon$-scalar contributions to the $\overline{\text{DRED}}$ correspondent diagrams, one recovers the results from CDR \textbf{diagram by diagram}.

\begin{table}[h!]
\begin{equation}
\begin{array}{|c|c|c|c|c|}
\hline
\text{Diagram}  & \text{A}_{\epsilon_{\text{scalar}}} & \text{B}_{\epsilon_{\text{scalar}}} 
\\ \hline 
c_{1} + c_{2} & - \frac{3}{4\epsilon} & - \frac{3}{4\epsilon}
\\ \hline 
f_{1} + f_{2} & - \frac{3}{4\epsilon} & - \frac{3}{4\epsilon}
\\ \hline 
h_{1} + h_{2} & \frac{3}{2\epsilon} & \frac{3}{2\epsilon}
\\ \hline 
\text{Sum} &  0 & 0 
\\ \hline  
\end{array}\nonumber
\end{equation}
\caption{Results for pure Yang-mills regarding $\epsilon$-scalar contributions}
\label{tab:epsilon}
\end{table}

\newpage

\subsection{QCD}

\begin{figure}[h!]
\centering
\subcaptionbox{}{\includegraphics[width=0.20\textwidth]{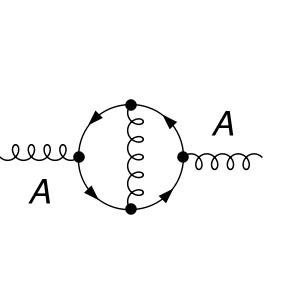}}%
\hfill
\subcaptionbox{}{\includegraphics[width=0.20\textwidth]{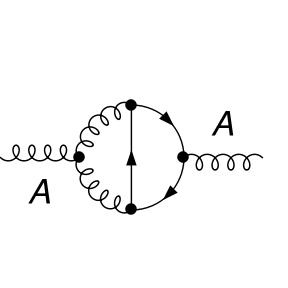}}%
\hfill
\subcaptionbox{}{\includegraphics[width=0.20\textwidth]{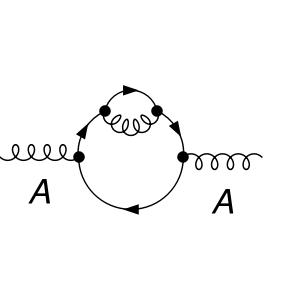}}%
\hfill
\subcaptionbox{}{\includegraphics[width=0.20\textwidth]{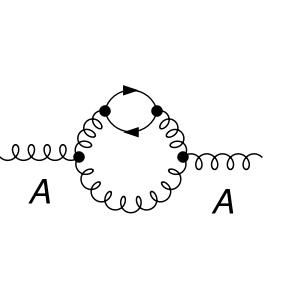}}%
\caption{Two-loop correction to the two-point function of the background field - fermionic contribution of QCD}
\label{fig:QCD}
\end{figure}

As our last example, we consider QCD. Since it is just a SU(3) Yang-Mills theory appended with $n_{f}$ flavors of fermions, we can reassess the results we obtained for the general Yang-Mills, specialize to SU(3) and include the corrections due to fermions, which are depicted in fig. \ref{fig:QCD} 

Adopting a similar convention of eq. \ref{eq: def A B},
\begin{equation}
\frac{i g_{s}^{4}n_{f}}{(4\pi)^{4}} \left[A g_{\mu\nu}p^{2} - B p_{\mu}p_{\nu}\right],
\end{equation}
\noindent
where $n_{f}$ is the number of fermions, we can express our results in tables \ref{tab: QCD ireg} and \ref{tab: QCD dreg}. The same patterns presented in the pure Yang-Mills theory appear here also, namely, the correlations among the coefficients of $\rho_{IREG}$; $\rho$ and $I_{log}^{2}(\lambda^{2})$,$I_{log}^{(2)}(\lambda^{2})$; $\epsilon^{-2}$. Also, as there is no realisation of topology T5 in this case, only terms proportional to $I_{log}(\lambda^{2})$ survive in the end result, as we already noticed in the QED case.

\begin{table}[h!]
\begin{equation}
\begin{array}{|c|c|c|c|c|c|c|c|c|}
\hline
\multicolumn{1}{ |c }{\multirow{2}{*}{\text{Diagram}}}  & \multicolumn{4}{|c}{\text{A}} & \multicolumn{4}{|c|}{\text{B}} 
\\ \cline{2-9}
& I_{log}^{(2)}(\lambda^{2}) & I_{log}^{2}(\lambda^{2}) & \rho_{IREG} & I_{log}(\lambda^{2}) & I_{log}^{(2)}(\lambda^{2}) & I_{log}^{2}(\lambda^{2}) & \rho_{IREG} & I_{log}(\lambda^{2}) \\ \hline 
a & -\frac{2}{9b} & \frac{2}{9b^2} & -\frac{2}{9b} & \frac{35}{27b} & -\frac{2}{9b} & \frac{2}{9b^2} & -\frac{2}{9b} & \frac{32}{27b}
\\ \hline 
b & -\frac{8}{b} & \frac{8}{b^2} & -\frac{8}{b} & \frac{44}{3b} &  -\frac{8}{b} & \frac{8}{b^2} & -\frac{8}{b} & \frac{50}{3b}
\\ \hline 
c & -\frac{16}{9b} & \frac{16}{9b^2} & -\frac{16}{9b} & \frac{208}{27b} &  -\frac{16}{9b} & \frac{16}{9b^2} & -\frac{16}{9b} & \frac{184}{27b}
\\ \hline 
d & \frac{10}{b} & -\frac{10}{b^2} & \frac{10}{b} & -\frac{97}{3b} &  \frac{10}{b} & -\frac{10}{b^2} & \frac{10}{b} & -\frac{100}{3b}
\\ \hline
\text{Sum} & 0 & 0 & 0 & -\frac{26}{3b} & 0 & 0 & 0 & -\frac{26}{3b}
\\ \hline  
\end{array}\nonumber
\end{equation}
\caption{Results for fermionic part of QCD using IREG where $\rho_{IREG} = I_{log}(\lambda^{2})\ln\left[-\frac{p^2}{\lambda^2}\right]$}
\label{tab: QCD ireg}
\end{table}

\begin{table}
\begin{equation}
\begin{array}{|c|c|c|c|c|}
\hline
\text{Diagram}  & \text{A}_{\text{CDR}} & \text{A}_{\text{DRED}}-\text{A}_{\text{CDR}} &\text{B}_{\text{CDR}} &  \text{B}_{\text{DRED}}-\text{B}_{\text{CDR}} 
\\ \hline 
a & \frac{1}{9\epsilon^{2}} + \frac{11-4\rho}{18\epsilon} & \frac{1}{9\epsilon} & \frac{1}{9\epsilon^{2}} + \frac{5-2\rho}{9\epsilon} & \frac{1}{9\epsilon}
\\ \hline 
b & \frac{4}{\epsilon^{2}} + \frac{13-8\rho}{\epsilon} & -\frac{2}{\epsilon}  & \frac{4}{\epsilon^{2}} + \frac{2(7-4\rho)}{\epsilon} & -\frac{2}{\epsilon}
\\ \hline 
c & \frac{8}{9\epsilon^{2}} + \frac{16(2-\rho)}{9\epsilon} & \frac{8}{9\epsilon}  & \frac{8}{9\epsilon^{2}} + \frac{4(7-4\rho)}{9\epsilon} & \frac{8}{9\epsilon}
\\ \hline 
d & -\frac{5}{\epsilon^{2}} - \frac{39-20\rho}{2\epsilon} & \frac{1}{\epsilon}  & -\frac{5}{\epsilon^{2}} - \frac{10(2-\rho)}{\epsilon} & \frac{1}{\epsilon}
\\ \hline 
\text{Sum} & -\frac{7}{3\epsilon} & 0 & -\frac{7}{3\epsilon} & 0
\\ \hline  
\end{array}\nonumber
\end{equation}
\caption{Results for fermionic part of QCD using CDR and $\overline{\text{DRED}}$}
\label{tab: QCD dreg}
\end{table}

One may notice that, in all diagrams, there is a mismatch between CDR and $\overline{\text{DRED}}$, although the sum is the same in both methods, as in all other examples we considered here. Finally, as in the case of YM, the result is gauge invariant and local, although we still need to add counterterms. The results of the counterterms follow a similar pattern of the one seen in the Yang-Mills theory, and they can be read from tables \ref{tab: ct QCD ireg}, and \ref{tab: ct QCD dreg}.	Finally, as in the pure Yang-Mills case, we reproduce previous results in the literature \cite{BFM:fermion}.

\begin{table}[h!]
\begin{equation}
\begin{array}{|c|c|c|c|c|c|c|c|c|}
\hline
\multicolumn{1}{ |c }{\multirow{2}{*}{\text{Counterterm}}}  & \multicolumn{4}{|c}{\text{A}} & \multicolumn{4}{|c|}{\text{B}} 
\\ \cline{2-9}
& I_{log}^{(2)}(\lambda^{2}) & I_{log}^{2}(\lambda^{2}) & \rho_{IREG} & I_{log}(\lambda^{2}) & I_{log}^{(2)}(\lambda^{2}) & I_{log}^{2}(\lambda^{2}) & \rho_{IREG} & I_{log}(\lambda^{2}) \\ \hline 
AA\hat{A} \text{ Coupling} & 0 & \frac{10}{3b^2} & -\frac{10}{3b} & \frac{56}{9b} &  0 & \frac{10}{3b^2} & -\frac{10}{3b} & \frac{56}{9b}
\\ \hline 
\text{Gluon self-energy} & 0 & -\frac{10}{3b^2} & \frac{10}{3b} & -\frac{92}{9b} &  0 & -\frac{10}{3b^2} & \frac{10}{3b} & -\frac{92}{9b}
\\ \hline 
\text{Sum} & 0 & 0 & 0 & -\frac{4}{b} & 0 & 0 & 0 & -\frac{4}{b}
\\ \hline  
\end{array}\nonumber
\end{equation}
\caption{Results for fermionic part of QCD using IREG where $\rho_{IREG} = I_{log}(\lambda^{2})\ln\left[-\frac{p^2}{\lambda^2}\right]$}
\label{tab: ct QCD ireg}
\end{table}

\begin{table}
\begin{equation}
\begin{array}{|c|c|c|c|c|}
\hline
\text{Counterterm}  & \text{A}_{\text{CDR}} & \text{A}_{\text{DRED}}-\text{A}_{\text{CDR}} &\text{B}_{\text{CDR}} &  \text{B}_{\text{DRED}}-\text{B}_{\text{CDR}} 
\\ \hline 
AA\hat{A} \text{ Coupling} & \frac{10}{3\epsilon^{2}} + \frac{2(28-15)\rho}{9\epsilon} & 0  & \frac{10}{3\epsilon^{2}} + \frac{2(28-15\rho)}{9\epsilon} & 0
\\ \hline 
\text{Gluon self-energy} & -\frac{10}{3\epsilon^{2}} - \frac{2(46-15\rho)}{9\epsilon} & 0  & -\frac{10}{3\epsilon^{2}} - \frac{2(46-15\rho)}{9\epsilon} & 0
\\ \hline 
\text{Sum} & -\frac{4}{\epsilon} & 0 & -\frac{4}{\epsilon} & 0
\\ \hline  
\end{array}\nonumber
\end{equation}
\caption{Results for fermionic part of QCD using CDR and $\overline{\text{DRED}}$}
\label{tab: ct QCD dreg}
\end{table}

\newpage

\subsection{Summary of the results}

In this subsection we collect the results we found, aiming to compute $Z_{A}$, the renormalisation function of the external gauge boson (the photon for QED, the background gluon field for Yang-Mills and QCD). Defining
\begin{align}
\label{eq:ZA pi}
Z_{A} = 1 + \frac{g^{2}}{(4\pi)^{2}} Z_{A}^{(1)} +  \frac{g^{4}}{(4\pi)^{4}}Z_{A}^{(2)}, 
\end{align}
one obtains for (scalar and spinorial) QED 
\begin{align}
\label{eq:ZA1}
Z_{A}^{(1)}|_{\text{IREG}}&=-\frac{4}{3b}I_{log}(\lambda^{2}),\quad Z_{A}^{(1)}|_{\text{CDR}}=-\frac{4}{3\epsilon},\\
Z_{A}^{(2)}|_{\text{IREG}}&= - \frac{4}{b}I_{log}(\lambda^{2}),\quad\quad Z_{A}^{(2)}|_{\text{CDR}}=-\frac{2}{\epsilon},\\
\end{align}
for SU(N) Yang-Mills
\begin{align}
\label{eq:ZA11}
Z_{A}^{(1)}|_{\text{IREG}}&=\frac{11}{3b}C_{A}I_{log}(\lambda^{2}),\quad Z_{A}^{(1)}|_{\text{CDR}}=\frac{11}{3\epsilon}C_{A},\\
Z_{A}^{(2)}|_{\text{IREG}}&=\frac{6}{b^{2}}C_{A}^{2}\left[I_{log}^{2}(\lambda^{2})-2bI_{log}^{(2)}(\lambda^{2})\right]+
\frac{7}{3b}C_{A}^{2}I_{log}(\lambda^{2}),\\ Z_{A}^{(2)}|_{\text{CDR}}&=\frac{17}{3\epsilon}C_{A}^{2},
\end{align}
and for QCD
\begin{align}
Z_{A}^{(1)}|_{\text{IREG}}&=\left(\frac{11}{b}-\frac{2}{3b}n_{f}\right)I_{log}(\lambda^{2}),\quad Z_{A}^{(1)}|_{\text{CDR}}=\frac{11}{\epsilon}-\frac{2}{3\epsilon}n_{f},\\
Z_{A}^{(2)}|_{\text{IREG}}&=\frac{54}{b^{2}}\left[I_{log}^{2}(\lambda^{2})-2bI_{log}^{(2)}(\lambda^{2})\right]+
\left(\frac{210}{b}-\frac{38}{3b}n_{f}\right)I_{log}(\lambda^{2}),\\ Z_{A}^{(2)}|_{\text{CDR}}&=\frac{51}{\epsilon}-\frac{19}{3\epsilon}n_{f}.
\label{eq:ZA2}
\end{align}
For completeness we have also computed and included the one-loop corrections, and we are already specializing to SU(3) when writing the QCD results. Notice that in the one-loop contribution, $I_{log}(\lambda^{2})/b$ and $\epsilon^{-1}$ share the same coefficient, as discussed in subsection \ref{sec:correlation}.

%\begin{equation}
%\mathcal{A_{\mu\nu}}=\left\{\frac{g_{s}^{2}}{(4\pi)^{2}}\left(\frac{11}{b}-\frac{2}{3b}n_{f}\right)I_{log}(\lambda^{2})+\frac{g_{s}^{4}}{(4\pi)^{4}}\left(\frac{102}{b}-\frac{38}{3b}n_{f}\right)I_{log}(\lambda^{2})\right\}i\left[ g_{\mu\nu}p^{2} - p_{\mu}p_{\nu}\right].
%\end{equation}
%\begin{equation}
%\mathcal{A_{\mu\nu}}=\left\{\frac{g_{e}^{2}}{(4\pi)^{2}}\left(\frac{11}{\epsilon}-\frac{2}{3\epsilon}n_{f}\right)+\frac{g_{e}^{4}}{(4\pi)^{4}}\left(\frac{51}{\epsilon}-\frac{19}{3\epsilon}\right)\right\}i\left[ g_{\mu\nu}p^{2} - p_{\mu}p_{\nu}\right].
%\end{equation}

\subsection{The $\beta$ function}

The $\beta$ function is obtained by adopting standard procedures exemplified in textbooks. However, in order to make the connection between the different regularisation methods clearer, we provide some details of the calculation. As usual, the $\beta$ function is defined by
\begin{equation}
\beta = \lambda \frac{\partial}{\partial \lambda} g_{R},
\label{eq:beta}
\end{equation} 
where $g_{R}$ is the (renormalized) gauge coupling of the theory considered, and $\lambda$ is the renormalisation scale (in dimensional methods, this is identified as $\mu_{DR}$). Until this point, no distinction between regularisations in fixed dimension and dimensional methods was done. To proceed further, we will adopt the framework of dimensional regularisation techniques, in which the (renormalized) coupling is replaced by
\begin{equation}
g_{R} = \mu^{\frac{4-d}{2}}\tilde{g}_{R}\quad\Rightarrow\quad \beta = \mu \frac{\partial}{\partial \mu} g_{R} = \frac{4-d}{2}g_{R} + \mu^{\frac{6-d}{2}}\frac{\partial}{\partial \mu}\tilde{g}_{R},
\end{equation}
where $\tilde{g}_{R}$ is an adimensional (renormalized) coupling. Notice that the equation above reduces to eq. (\ref{eq:beta}) when using methods in fixed dimension. One may also introduce $Z_{g}$ which is the renormalisation constant related to the coupling $g$ satisfying $g_{0} = Z_{g} g_{R}$, where $g_{0}$ is the bare coupling. Therefore, the $\beta$-function can also be given by the related equation in general
\begin{equation}
\beta = -g_{R} \lambda \frac{\partial}{\partial \lambda} \ln Z_{g}.
\label{eq:betaZ}
\end{equation} 
In the background field method, the relation $Z_{g}=Z_{A}^{-1/2}$ is valid, which reduces the calculation of the $\beta$ function in a non-abelian theory to the knowledge of \textbf{only} two-point functions. To proceed, we assume that $Z_{A}$ can be expanded in the (renormalized) adimensional coupling constant $\tilde{g}_{R}$
\begin{equation}
Z_{A} = 1 + A_{1} \tilde{g}_{R}^{2} + A_{2} \tilde{g}_{R}^{4},
\label{eq:ZA}
\end{equation}  
where $A_{i}$ is related to the counterterm of the $i$-order that renormalize the $i$-loop correction to the two-point function in the background field $A$. This amounts to 
\begin{equation}
\beta = \frac{g_{R}}{2} \lambda \frac{\partial}{\partial \lambda} \left[A_{1}\tilde{g}_{R}^{2} + \left(A_{2} - \frac{A_{1}^{2}}{2}\right)\tilde{g}_{R}^{4}\right].
\end{equation}
Notice that the above formula is valid for methods in fixed dimension as well, since for those $\tilde{g}_{R} = g_{R}$.  To proceed further, one has to choose a subtraction scheme, which will be the MS-subtraction scheme (for methods in fixed dimension we will discuss this point later). Therefore, all $A_{i}$ will be independent of $\mu$ which implies 
\begin{align}
\beta &= \frac{g_{R}}{2} \left[A_{1}\mu \frac{\partial}{\partial \mu}(\tilde{g}_{R}^{2}) + \left(A_{2} - \frac{A_{1}^{2}}{2}\right)\mu \frac{\partial}{\partial \mu}(\tilde{g}_{R}^{4})\right],\nonumber\\
&= \frac{\tilde{g}_{R}}{2} \left[2\; A_{1} \tilde{g}_{R} \left(\beta-\frac{4-d}{2}g_{R}\right) + 4\tilde{g}_{R}^{3}\left(A_{2} - \frac{A_{1}^{2}}{2}\right)\left(\beta - \frac{4-d}{2}g_{R}\right)\right],\nonumber\\
&=  \frac{d-4}{2}g_{R}\left[ A_{1} \tilde{g}_{R}^{2} + 2 A_{2} \tilde{g}_{R}^{4}\right].
\label{eq:beta_dred}
\end{align}
As standard, one can also define the expansion of the $\beta$-function in the adimensional coupling constant as
\begin{equation}
\beta = -g_{R}\left[\beta_{0} \left(\frac{\tilde{g}_{R}}{4\pi}\right)^{2} + \beta_{1} \left(\frac{\tilde{g}_{R}}{4\pi}\right)^{4}\right];
\label{eq:beta pi}
\end{equation}  
which, after careful comparison between eqs. \ref{eq:beta pi}, \ref{eq:ZA pi} and \ref{eq:ZA}, allows the identification
\begin{equation}
\beta_{0} = \epsilon Z_{A}^{(1)}, \quad \beta_{1} = 2 \epsilon Z_{A}^{(2)},
\end{equation}
since $d=4-2\epsilon$. Notice that, since the $\beta$ function is finite, even the two-loop coefficient of $Z_{A}$ must only have terms up to $\epsilon^{-1}$. This can be explicitly confirmed by looking at eqs. \ref{eq:ZA1} to \ref{eq:ZA2}.

Regarding methods in fixed dimension, there are some differences among them in the definition of the counterterms. In FDR \cite{FDR} as well as DIFR \cite{FREEDMAN}, the divergent expressions are replaced by finite ones, meaning that divergences are automatically removed by applying the method. In IREG, divergences are kept, being identified as basic divergent integrals such as $I_{log}^{(2)}(\lambda^{2})$, and $I_{log}(\lambda^{2})$. As can be immediately seen, in IREG the divergent part will depend on the renormalisation scale $\lambda$, while in dimensional methods this does not occur. By defining the MS-subtraction scheme in IREG as the removal of only basic divergent integrals, the IREG version of eq.(\ref{eq:beta_dred}) will be 
\begin{align}
\beta &= \frac{g_{R}}{2} \left[g_{R}^{2}\lambda \frac{\partial}{\partial \lambda}(A_{1})+A_{1}\lambda \frac{\partial}{\partial \lambda}(g_{R}^{2}) + g_{R}^{4}\lambda \frac{\partial}{\partial \lambda}\left(A_{2} - \frac{A_{1}^{2}}{2}\right)+ \left(A_{2} - \frac{A_{1}^{2}}{2}\right)\lambda \frac{\partial}{\partial \lambda}(g_{R}^{4})\right],\nonumber\\
&= \frac{g_{R}}{2} \left[g_{R}^{2}\lambda \frac{\partial}{\partial \lambda}(A_{1})+2\; A_{1} g_{R} \beta + g_{R}^{4}\lambda \frac{\partial}{\partial \lambda}\left(A_{2} - \frac{A_{1}^{2}}{2}\right) + 4g_{R}^{3}\left(A_{2} - \frac{A_{1}^{2}}{2}\right)\beta\right],\nonumber\\
&=  -g_{R}\left[ -\frac{g_{R}^{2}}{2}\lambda \frac{\partial}{\partial \lambda}A_{1}   -\frac{g_{R}^{4}}{2}\lambda \frac{\partial}{\partial \lambda}A_{2}\right],
\label{eq:beta_ireg}
\end{align}
which, by comparing eqs. \ref{eq:beta pi}, \ref{eq:ZA pi} and \ref{eq:ZA}, implies
\begin{equation}
\beta_{0} = - \frac{1}{2}\lambda \frac{\partial}{\partial \lambda}Z_{A}^{(1)}, \quad \beta_{1} = - \frac{1}{2} \lambda \frac{\partial}{\partial \lambda}Z_{A}^{(2)}
\label{eq:betai}
\end{equation}
As already pointed out, the $Z_{A}^{(i)}$ in IREG will depend on basic divergent integrals, which implies that the derivatives of those with respect to $\lambda$ will be needed. They can be obtained in a straightforward way as shown in eqs. \ref{gerder}. For our purposes here, we only need
\begin{equation}
\lambda \frac{\partial}{\partial \lambda}I_{log}(\lambda^{2}) = -2b; \quad
\lambda \frac{\partial}{\partial \lambda}I_{log}^{(2)}(\lambda^{2}) = -2b - 2I_{log}(\lambda^{2}); \quad
%\lambda \frac{\partial}{\partial \lambda}I_{log}^{2}(\lambda^{2}) = -4bI_{log}(\lambda^{2}).
\label{eq:dev}
\end{equation}
Notice that the second of the equations above is still divergent. However, the combination $I_{log}^{2}(\lambda^{2})-2bI_{log}^{(2)}(\lambda^{2})$, which appears in 
$Z_{A}^{(2)}|_{\text{IREG}}$, will give a finite result as it should. Finally, by applying our results collected in eqs. \ref{eq:ZA1} to \ref{eq:ZA2}, we obtain the well-known one and two-loop contributions for the gauge $\beta$ coupling in QED (scalar and spinorial) \cite{QED:3loop}
\begin{align}
\beta_{0}|_{\text{IREG}} = -\frac{4}{3}; \quad &\beta_{0}|_{\text{CDR}} = -\frac{4}{3}; \\
\beta_{1}|_{\text{IREG}} = -4; \quad &\beta_{1}|_{\text{CDR}} = -4;
\end{align}
for pure Yang-Mills \cite{Abbott}
\begin{align}
\beta_{0}|_{\text{IREG}} = \frac{11}{3}C_{A}; \quad &\beta_{0}|_{\text{CDR}} = \frac{11}{3}C_{A}; \\
\beta_{1}|_{\text{IREG}} = \frac{34}{3}C_{A}^{2}; \quad &\beta_{1}|_{\text{CDR}} = \frac{34}{3}C_{A}^{2};
\end{align}
and QCD \cite{BFM:fermion,QCD:2loop}
\begin{align}
\beta_{0}|_{\text{IREG}} = 11-\frac{2}{3}n_{f}; \quad &\beta_{0}|_{\text{CDR}} = 11-\frac{2}{3}n_{f}; \\
\beta_{1}|_{\text{IREG}} = 102-\frac{38}{3}n_{f}; \quad &\beta_{1}|_{\text{CDR}} = 102-\frac{38}{3}n_{f}
\end{align}

As can be readily seen, the results of all regularization methods applied in this work agree. We emphasize that this was expected since in all cases we are adopting a subtraction scheme independent of the mass which, in dimensional methods,  translates in the removal only of poles in $\epsilon$ while in IREG it amounts to the subtraction of BDI's. Therefore, for the methods under study in this contribution the first two coefficients of the $\beta$-function of gauge couplings are universal \cite{TARRACH}. 

\section{Concluding remarks}
\label{sec:conclusions}

To extract any deviation between theory and experimental data in the SM  as well as test BSM theories, precision observables demand at least $N^2LO$ and $N^3LO$ approximations involving multi-loop Feynman diagrams. Clearly the choice of the regularisation scheme to separate UV and IR divergencies of multi-loop amplitudes that enter into a computer code is guided by consistency and expediency.  For the reasons we have discussed in the introduction, practical and  symmetry-preserving regularisation frameworks that work fully in the physical dimension are desirable especially when dealing with dimensional-specific models in which the analytical continuation in the space-time dimension is ambiguous. This has justified to exploit quasi-dimensional methods such as DRED and FDH. They have been successfully employed in calculations in gauge and supersymmetric models after having their consistency validated, order by order in perturbation theory, through verification of Ward identities via quantum action principles. The main drawback of such schemes is that some modifications at Lagrangian level become necessary. For instance  higher covariant derivative terms  improve the ultraviolet behaviour of the propagators at the expense of  complicating the Feynman rules (for recent application of this technique in the context of supersymmetric theories see \cite{HD}). In the case of DRED or FDH, evanescent scalar $\epsilon$-particles add a $\cal{L}_\epsilon$ term to QCD Lagrangian as a result of decomposing the quasi-4-dimensional gluon field. Moreover, two new coupling constants besides $g_s$ emerge as a result to the coupling of $\epsilon$-scalars to (anti-)quarks, namely $g_\epsilon$, and a quartic $\epsilon$-scalar coupling  $g_{4\epsilon}$, with their respective $\beta$-functions and anomalous dimensions. Whilst such modifications are crucial for inner consistency of the method as well as shedding light on the ultraviolet and infrared factorisation structure of the amplitudes, they are unnecessary in fully non-dimensional  methods such as IREG.

In other to raise a non-dimensional scheme such as IREG to the level of more conventional methods a series calculations had to be performed. Firstly, show that a  program that displays the UV (and IR) content of an amplitude as a BDI, without recoursing to explicit evaluation, can be consistently and invariantly extended beyond one loop respecting gauge invariance. We have explicitly verified that this is the case by fully evaluating the contributions to the $\beta$-function of abelian and non-abelian models. We have verified the conjecture (proved for the abelian case) that a constrained version of IREG that sets to zero well defined surface terms in consonance with momentum routing invariance in the loops of Feynman diagrams automatically implements gauge invariance. We have obtained the renormalization constants by conducting the subtraction of subdivergences within IREG and compared with CDR and DRED. It is well-known that CDR and DRED are not equivalent in general in the sense that the residues of the poles in $\epsilon$ do not coincide. In this respect it is noteworthy, as we have explicitly verified, that evaluating the BDI's of IREG in $4-2\epsilon$ dimensions in the end of the calculation does not yield the same residues for the poles of arbitrary orders. Nonetheless, as we have shown in tables 2 to 18, a systematic summation among different contributions from Feynman graphs and counterterms renders an identical result for CDR, DRED. %, and $\overline{\text{DRED}}$. 
Matter-of-factly a tuned cancellation of $\epsilon$-scalar contributions take place. Because IREG does not
recourse to such modifications, it would be interesting to perform a calculation where such cancellations do not occur in %$\overline{\text{DRED}}$
DRED such as in the $g + g \rightarrow q + \bar{q} + g$ \cite{BEENAKKER} or $H \rightarrow g + g$ \cite{BROGGIO} scatterings to $NLO$ and $N^2LO$. Finally, we have computed the universal two-loop $\beta$-functions of gauge coupling in scalar and spinorial QED as well as pure Yang-Mills and QCD in a fully quadridimensional framework by defining the renormalization constants as BDI's. Derivatives of BDI's  with respect to a renormalization scale that naturally appears through a scale relation are also expressable as BDI's. This enable us to perform the calculations without explicitly evaluating the BDI's.

In order to pursuit the IREG program to apply it to precision calculations, it is important  to show that IREG respects the factorisation properties of infrared divergences in QCD as well as to evaluate the cusp anomalous dimensions. This can be achieved in two ways: either by parametrising the infrared divergences in IREG as $\ln \mu^2$ as $\mu \rightarrow 0$ or by using a parametrisation of infrared divergences in the reciprocal space in terms of infrared BDI in the coordinate space. Both approaches are under active investigation.

\appendix
\section{Feynman rules}
\label{app:feynman}

\begin{figure}
	\begin{subfigure}[!h]{0.4\textwidth}
		\includegraphics[width=\textwidth]{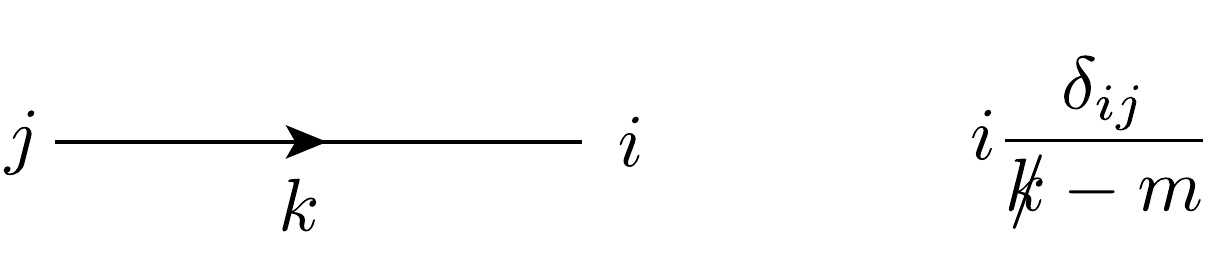}
	\end{subfigure}
\quad
	\begin{subfigure}[!h]{0.4\textwidth}
		\includegraphics[width=\textwidth]{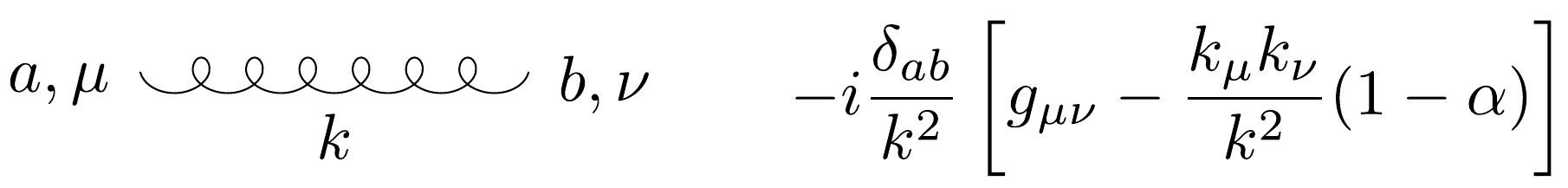}
	\end{subfigure}
\quad
    \begin{subfigure}[!h]{0.4\textwidth}
    	\includegraphics[width=\textwidth]{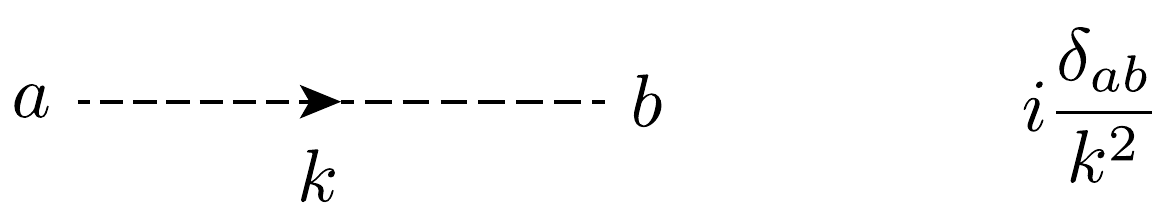}
    \end{subfigure}
\quad
\begin{subfigure}[!h]{0.4\textwidth}
	\includegraphics[width=\textwidth]{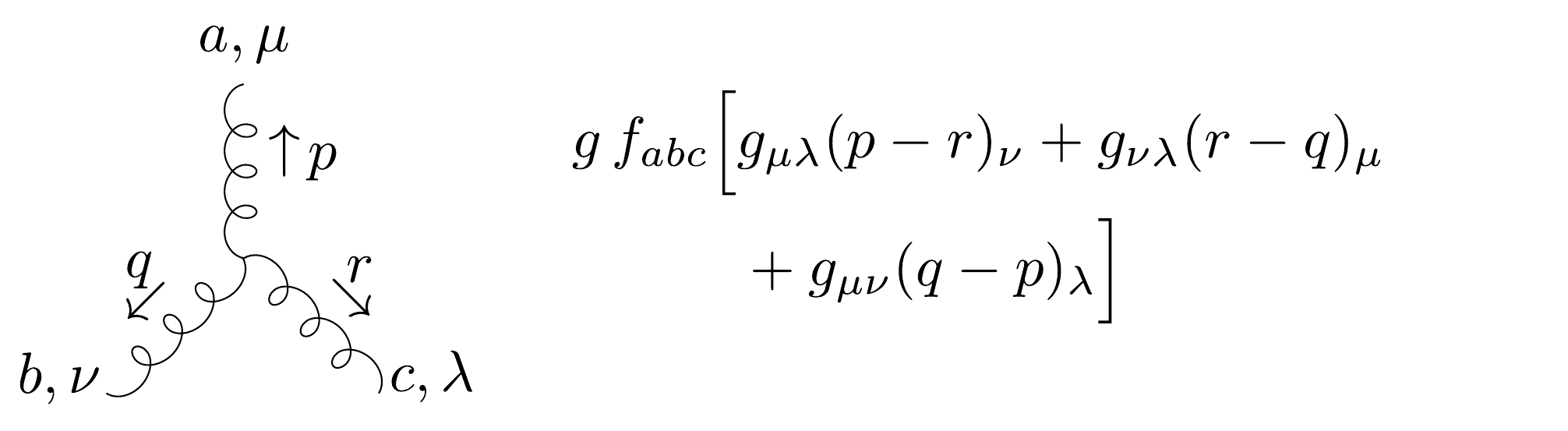}
\end{subfigure}
\quad
\begin{subfigure}[!h]{0.4\textwidth}
\includegraphics[width=\textwidth]{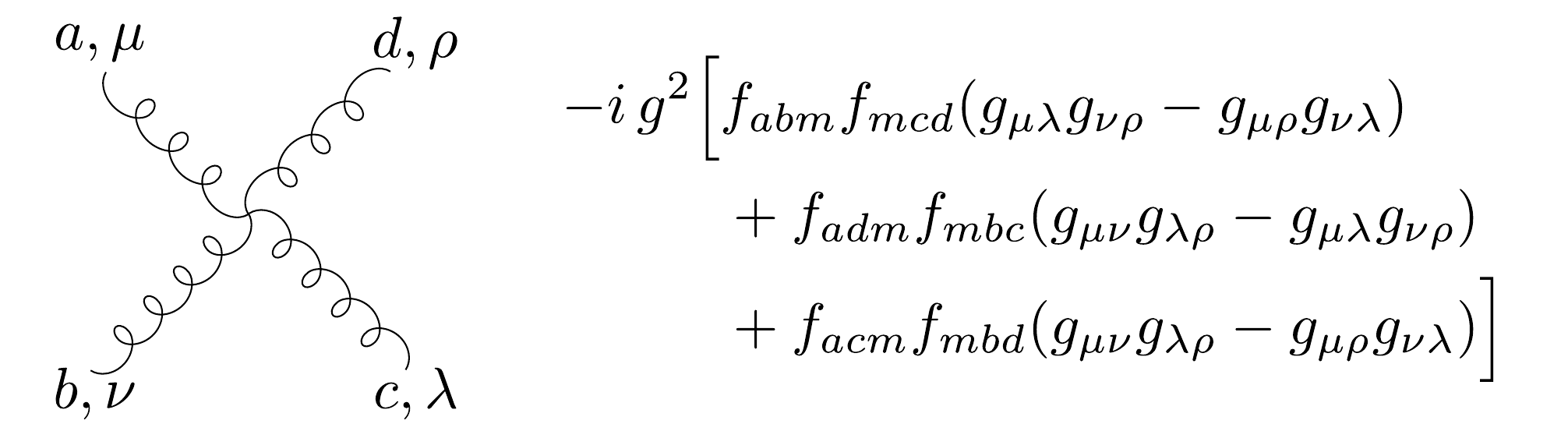}
\end{subfigure}
\quad
\begin{subfigure}[!h]{0.4\textwidth}
\includegraphics[width=\textwidth]{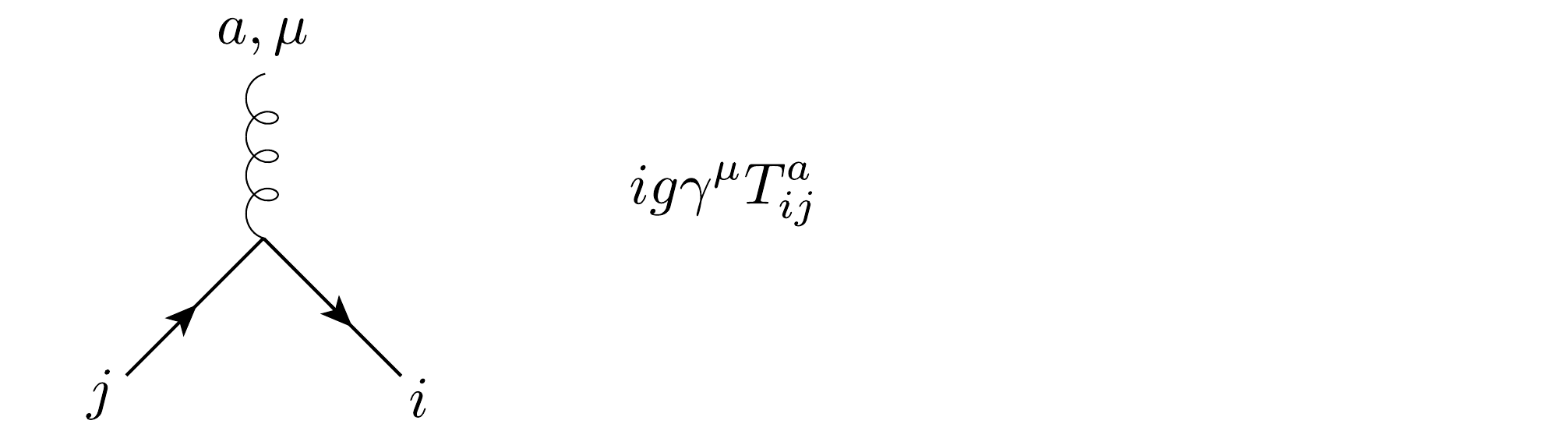}
\end{subfigure}
\quad
\begin{subfigure}[!h]{0.4\textwidth}
\includegraphics[width=\textwidth]{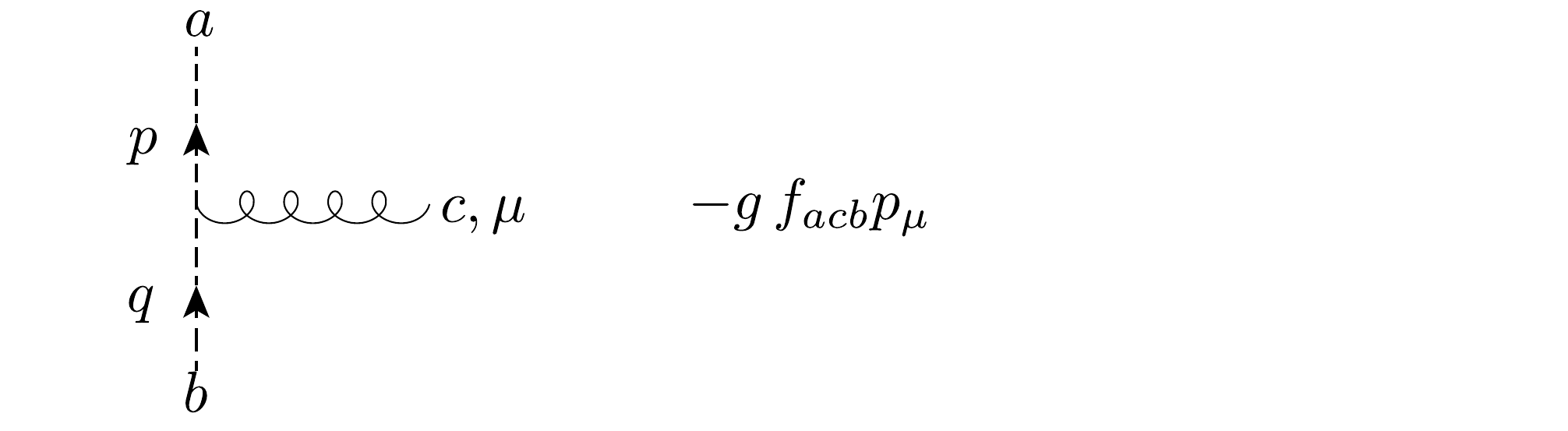}
\end{subfigure}
\quad
\begin{subfigure}[!h]{0.4\textwidth}
\includegraphics[width=\textwidth]{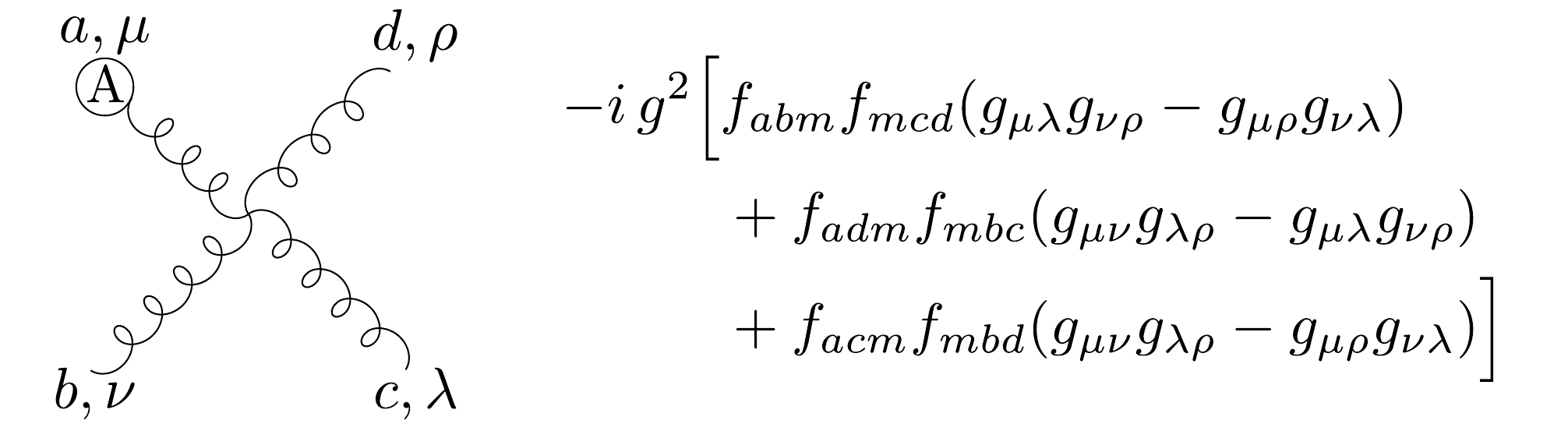}
\end{subfigure}
\quad
\begin{subfigure}[!h]{0.4\textwidth}
\includegraphics[width=\textwidth]{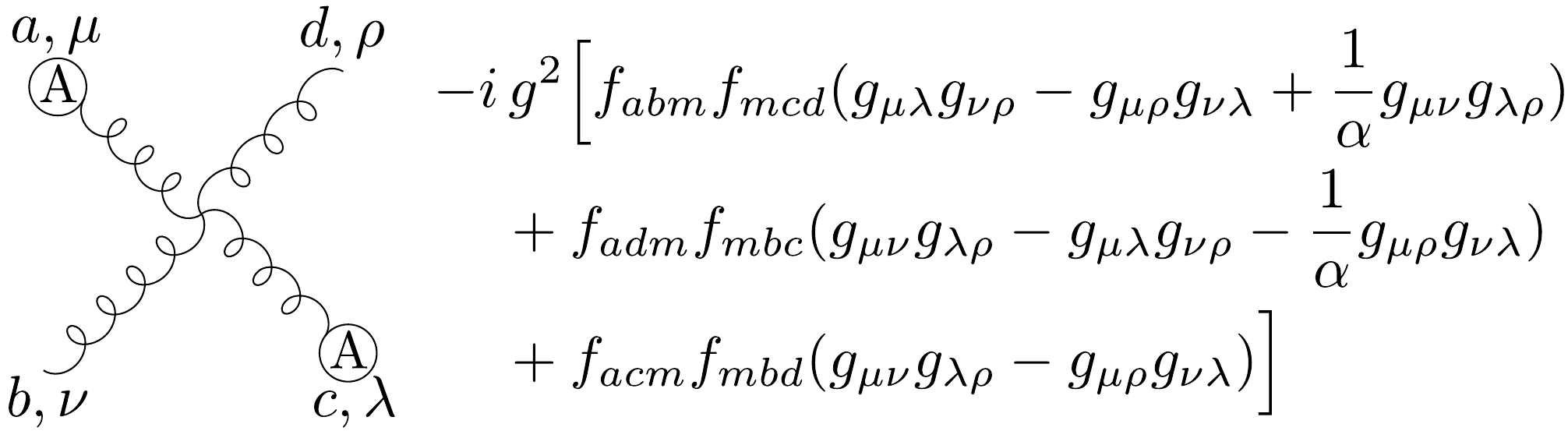}
\end{subfigure}
\quad
\begin{subfigure}[!h]{0.4\textwidth}
\includegraphics[width=\textwidth]{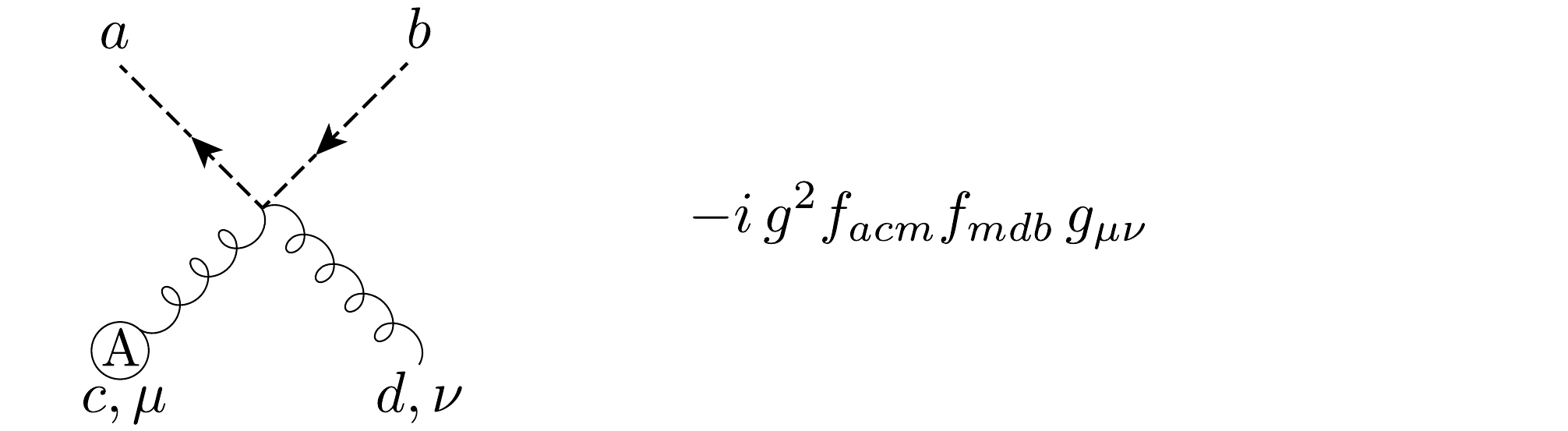}
\end{subfigure}
\quad
\begin{subfigure}[!h]{0.4\textwidth}
\includegraphics[width=\textwidth]{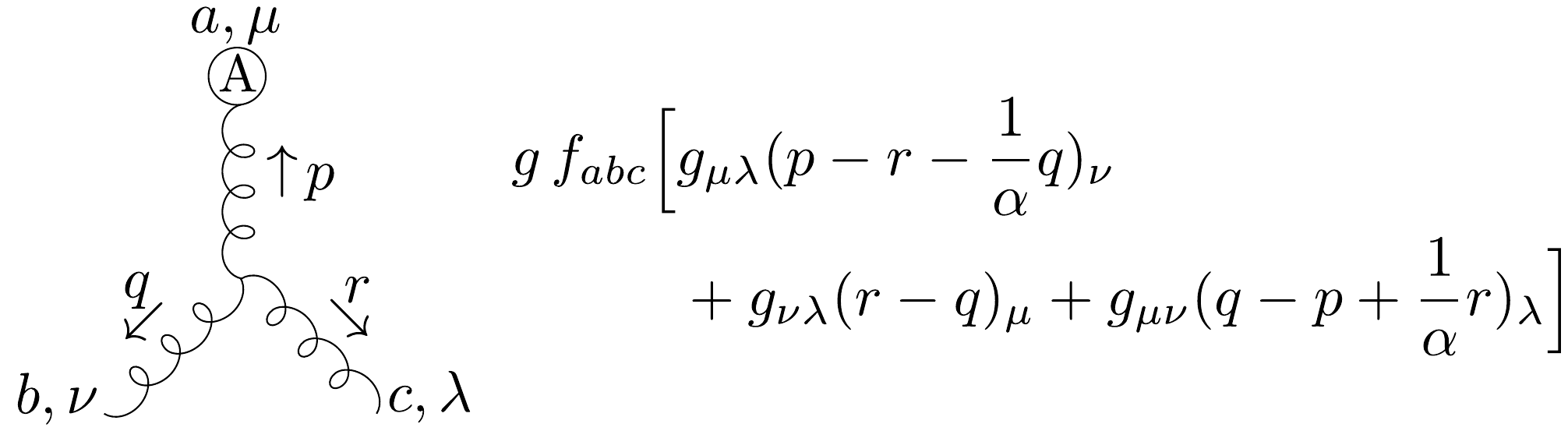}
\end{subfigure}
\quad
\begin{subfigure}[!h]{0.4\textwidth}
\includegraphics[width=\textwidth]{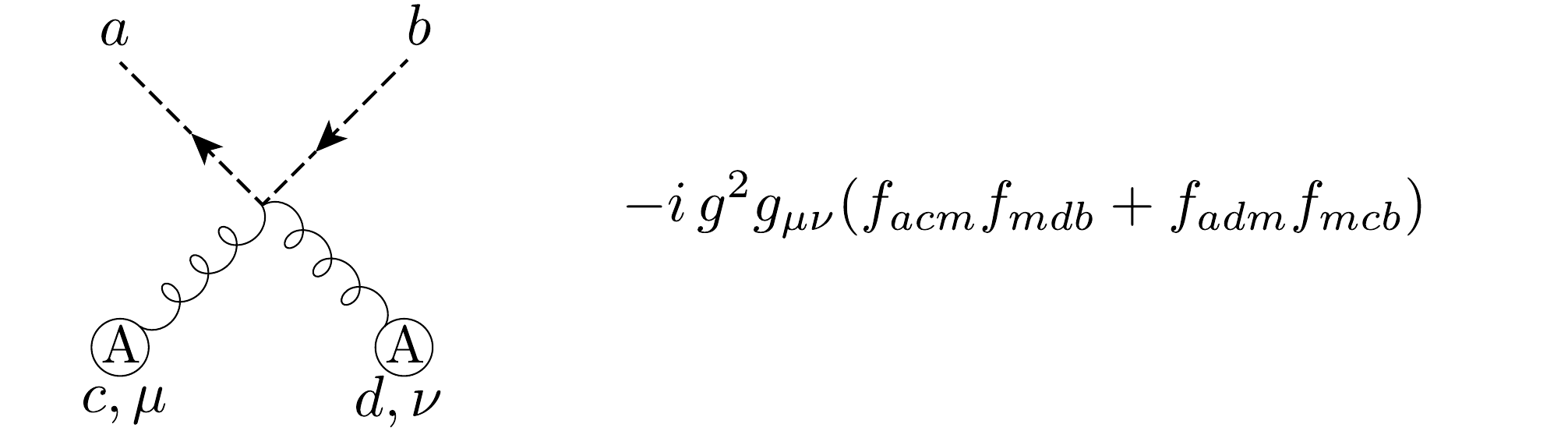}
\end{subfigure}
\quad
\begin{subfigure}[!h]{0.4\textwidth}
\includegraphics[width=\textwidth]{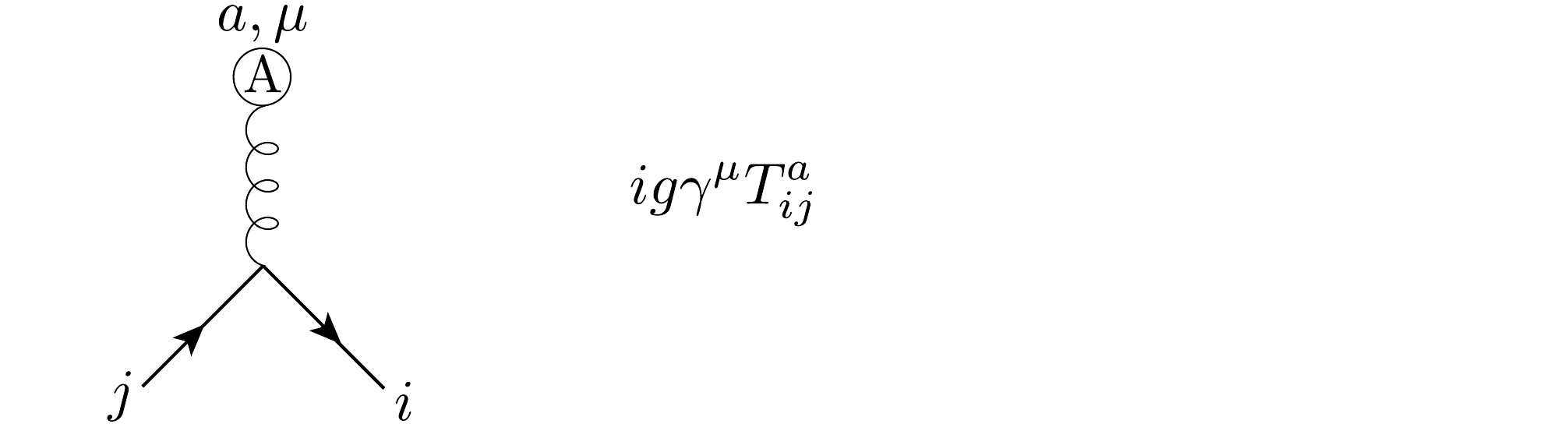}
\end{subfigure}
\quad \quad \quad \quad \quad \quad \quad \quad
\begin{subfigure}[!h]{0.4\textwidth}
\includegraphics[width=\textwidth]{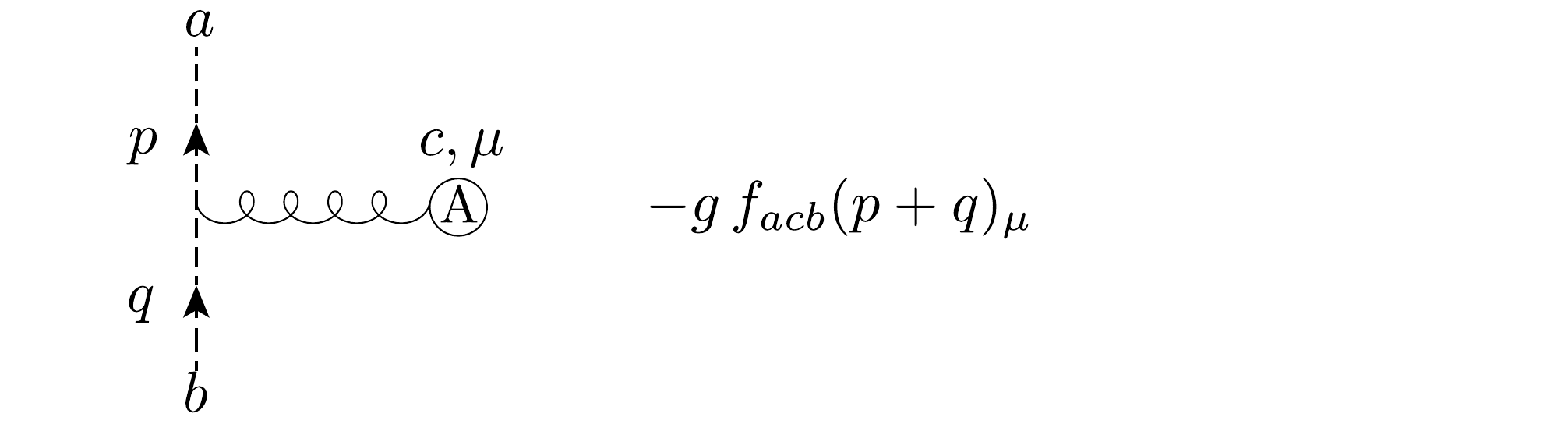}
\end{subfigure}
\caption{Feynman rules}
\label{fig:FR}
\end{figure}

For scalar and spinorial $QED$ we refer to \cite{Schwartz} for the Feynman rules and conventions. As for $QCD$ with a background field $A$ and gauge fixing parameter $\alpha$, we follow the conventions of \cite{Abbott} to yield the following rules in figure \ref{fig:FR}.

\section{Explicit results of integrals}
\label{app: integrals}

Here we present explicit results for the integrals used in this work. They were obtained combining in-house routines with 1-loop evaluation of some integrals performed with \textit{Package X}\cite{PACKAGEX}.

\subsection{1-loop: two point functions}

\begin{align}
\int_{k} \frac{1}{k^2(k+p)^2}&= I_{log}(\lambda^{2})- b \ln\left(-\frac{p^2}{\lambda^2}\right) + 2b,\\
\int_{k} \frac{k_{\mu}}{k^2(k+p)^2}&=-\frac{p_{\mu}}{2}\left[I_{log}(\lambda^{2})- b \ln\left(-\frac{p^2}{\lambda^2}\right) + 2b\right],\\
\label{app:1loop}
\int_{k} \frac{k_{\mu}k_{\nu}}{k^2(k+p)^2}&=-\frac{g_{\mu\nu}p^{2}}{12}\left[I_{log}(\lambda^{2})- b \ln\left(-\frac{p^2}{\lambda^2}\right) + \frac{8b}{3}\right]\nonumber\\
&\quad\quad+\frac{p_{\mu}p_{\nu}}{3}\left[I_{log}(\lambda^{2})- b \ln\left(-\frac{p^2}{\lambda^2}\right) + \frac{13b}{6}\right]
\end{align}

\subsection{1-loop: three point functions}

\begin{align}
\int_{k} \frac{k_{\mu}k_{\nu}}{k^4(k+p)^2}&=\frac{g_{\mu\nu}}{4}\left[I_{log}(\lambda^{2})- b \ln\left(-\frac{p^2}{\lambda^2}\right) + 2b\right] + \frac{p_{\mu}p_{\nu}}{p^2}\frac{b}{2}\\
\int_{k} \frac{k_{\mu}k_{\nu}k_{\rho}}{k^2(k+p_{1})^2(k+p_{2})^2}&=\left[g_{\mu\nu}(p_{1}+p_{2})_{\rho}+g_{\nu\rho}(p_{1}+p_{2})_{\mu}+g_{\mu\rho}(p_{1}+p_{2})_{\nu}\right]\left[-\frac{I_{log}(\lambda^{2})}{12}+\mbox{finite}\right]
\end{align}

\subsection{2-loop: explicit results involving logarithms}

We recall the general structure  given in eq. \ref{I}

\begin{align}
&I^{\nu_{1}\ldots \nu_{m}}\!=\!\!\int\limits_{k_{l}}\!\frac{A^{\nu_{1}\ldots \nu_{m}}(k_{l},q_{i})}{\prod_{i}[(k_{l}-q_{i})^{2}-\mu^{2}]}\ln^{l-1}\!\left(\!-\frac{k_{l}^{2}-\mu^{2}}{\lambda^{2}}\right)\!,
%\label{I}
\end{align}
Considering the 2-loop case ($l=2$), it simplifies to (after relabeling $k_{2}=k$) 
\begin{align}
&I^{\nu_{1}\ldots \nu_{m}}\!=\!\!\int\limits_{k}\!\frac{A^{\nu_{1}\ldots \nu_{m}}(k,q_{i})}{\prod_{i}[(k-q_{i})^{2}-\mu^{2}]}\ln\!\left(\!-\frac{k^{2}-\mu^{2}}{\lambda^{2}}\right)\!,
%\label{I}
\end{align}

\subsubsection{One point functions}

\begin{align}
\int_{k} \frac{1}{(k+p)^2}\ln\left(-\frac{k^2}{\lambda^2}\right)&= \frac{p^{2}}{2}I_{log}(\lambda^{2}) +\mbox{finite},
\end{align}

\subsubsection{Two point functions}

\begin{align}
\int_{k} \frac{1}{k^2(k+p)^2}\ln\left(-\frac{k^2}{\lambda^2}\right)&= I_{log}^{(2)}(\lambda^{2}) +\mbox{finite},\\
\int_{k} \frac{k_{\mu}}{k^2(k+p)^2}\ln\left(-\frac{k^2}{\lambda^2}\right)&=-\frac{p_{\mu}}{2}\left[I_{log}^{(2)}(\lambda^{2})+\frac{I_{log}(\lambda^{2})}{2} +\mbox{finite}\right],\\
\int_{k} \frac{k_{\mu}k_{\nu}}{k^2(k+p)^2}\ln\left(-\frac{k^2}{\lambda^2}\right)&=\frac{g_{\mu\nu}}{12}\left[-I_{log}^{(2)}(\lambda^{2}) + \frac{I_{log}(\lambda^{2})}{6} + \mbox{finite}\right]\nonumber\\
&\quad\quad+\frac{p_{\mu}p_{\nu}}{3}\left[I_{log}^{(2)}(\lambda^{2})+\frac{5}{6}I_{log}(\lambda^{2})+\mbox{finite}\right],
\end{align}

\subsubsection{Three point functions}

\begin{align}
\int_{k} \frac{k_{\mu}k_{\nu}}{k^4(k+p)^2}\ln\left(-\frac{k^2}{\lambda^2}\right)&=\frac{g_{\mu\nu}}{4}\left[I_{log}^{(2)}(\lambda^{2}) +\frac{I_{log}(\lambda^{2})}{2} + \mbox{finite}\right]  + \frac{p_{\mu}p_{\nu}}{p^2}\mbox{finite},
\end{align}

\subsection{2-loop: overlapped integrals}

We will have the general structure
\begin{equation}
I[f(k,q)] = \int_{k} \frac{f(k,q)}{k^2(k-p)^2(k-q)^2 q^2 (q-p)^2}
\end{equation}
\begin{align}
I[k.q] &= bI_{log}(\lambda^{2})+\mbox{finite}\\
I[k_{\mu}q_{\nu}] &= \frac{g_{\mu\nu}}{4}\left[I_{log}(\lambda^{2})+\mbox{finite}\right]+\frac{p_{\mu}p_{\nu}}{p^2}\mbox{finite}\\
I[k_{\mu}k_{\nu}] &= \frac{g_{\mu\nu}}{4}\left\{I_{log}^{2}(\lambda^{2})-bI_{log}^{(2)}(\lambda^{2})+ b I_{log}(\lambda^{2})\left[\frac{9}{2}-\ln\left(-\frac{p^2}{\lambda^2}\right)\right]+\mbox{finite}\right\}\nonumber\\
&\quad+\frac{p_{\mu}p_{\nu}}{p^2}\mbox{finite}\\
I[k_{\mu}q_{\nu}q.p] &= \frac{p_{\mu}p_{\nu}}{8}\left\{I_{log}^{2}(\lambda^{2})-bI_{log}^{(2)}(\lambda^{2})+ b I_{log}(\lambda^{2})\left[\frac{11}{2}-\ln\left(-\frac{p^2}{\lambda^2}\right)\right]+\mbox{finite}\right\}\nonumber\\
&\quad+\frac{g_{\mu\nu}p^2}{8}\left[I_{log}(\lambda^{2})+\mbox{finite}\right]\\
I[k_{\mu}k_{\nu}q.p] &= \frac{g_{\mu\nu}}{8}\left\{I_{log}^{2}(\lambda^{2})-bI_{log}^{(2)}(\lambda^{2})+ b I_{log}(\lambda^{2})\left[\frac{9}{2}-\ln\left(-\frac{p^2}{\lambda^2}\right)\right]+\mbox{finite}\right\}\nonumber\\
&\quad+\frac{p_{\mu}p_{\nu}}{4}\left[I_{log}(\lambda^{2})+\mbox{finite}\right]\\
I[k_{\mu}k.q] &= \frac{p_{\mu}}{8}\left\{I_{log}^{2}(\lambda^{2})-bI_{log}^{(2)}(\lambda^{2})+ b I_{log}(\lambda^{2})\left[\frac{19}{2}-\ln\left(-\frac{p^2}{\lambda^2}\right)\right]+\mbox{finite}\right\}\\
I[k_{\mu}q_{\nu}k.q] &= \frac{g_{\mu\nu}}{24}\left\{-I_{log}^{2}(\lambda^{2})+bI_{log}^{(2)}(\lambda^{2})- b I_{log}(\lambda^{2})\left[\frac{29}{6}-\ln\left(-\frac{p^2}{\lambda^2}\right)\right]+\mbox{finite}\right\}\nonumber\\
&\quad+\frac{p_{\mu}p_{\nu}}{6}\left\{I_{log}^{2}(\lambda^{2})-bI_{log}^{(2)}(\lambda^{2})+ b I_{log}(\lambda^{2})\left[\frac{85}{12}-\ln\left(-\frac{p^2}{\lambda^2}\right)\right]+\mbox{finite}\right\}
\end{align}
\begin{align}
I[k^2] &= I_{log}^{2}(\lambda^{2})-bI_{log}^{(2)}(\lambda^{2})+ b I_{log}(\lambda^{2})\left[4-\ln\left(-\frac{p^2}{\lambda^2}\right)\right]+\mbox{finite}\\
I[k^2 k_{\mu}] &= \frac{p_{\mu}}{4}\left\{3I_{log}^{2}(\lambda^{2})-3bI_{log}^{(2)}(\lambda^{2})+ b I_{log}(\lambda^{2})\left[\frac{25}{2}-3\ln\left(-\frac{p^2}{\lambda^2}\right)\right]+\mbox{finite}\right\}\\
I[k^2 q_{\mu}] &= \frac{p_{\mu}}{2}\left\{I_{log}^{2}(\lambda^{2})-bI_{log}^{(2)}(\lambda^{2})+ b I_{log}(\lambda^{2})\left[\frac{9}{2}-\ln\left(-\frac{p^2}{\lambda^2}\right)\right]+\mbox{finite}\right\}\\
I[k^2 k_{\mu}q_{\nu}] &= \frac{g_{\mu\nu}}{24}\left\{-I_{log}^{2}(\lambda^{2})+bI_{log}^{(2)}(\lambda^{2})- b I_{log}(\lambda^{2})\left[\frac{29}{6}-\ln\left(-\frac{p^2}{\lambda^2}\right)\right]+\mbox{finite}\right\}\nonumber\\
&\quad+\frac{p_{\mu}p_{\nu}}{24}\left\{10I_{log}^{2}(\lambda^{2})-10bI_{log}^{(2)}(\lambda^{2})+ b I_{log}(\lambda^{2})\left[\frac{139}{3}-10\ln\left(-\frac{p^2}{\lambda^2}\right)\right]+\mbox{finite}\right\}\\
I[k^2 k.q] &= \frac{1}{4}\left[I_{log}^{2}(\lambda^{2})-bI_{log}^{(2)}(\lambda^{2})+ b I_{log}(\lambda^{2})\left[\frac{9}{2}-\ln\left(-\frac{p^2}{\lambda^2}\right)\right]+\mbox{finite}\right]\\
I[k^2 q^2] &=0\\
I[k^2 q_{\mu}q_{\nu}] &= \frac{g_{\mu\nu}}{12}\left\{-I_{log}^{2}(\lambda^{2})+bI_{log}^{(2)}(\lambda^{2})- b I_{log}(\lambda^{2})\left[\frac{29}{6}-\ln\left(-\frac{p^2}{\lambda^2}\right)\right]+\mbox{finite}\right\}\nonumber\\
&\quad+\frac{p_{\mu}p_{\nu}}{3}\left\{I_{log}^{2}(\lambda^{2})-bI_{log}^{(2)}(\lambda^{2})+ b I_{log}(\lambda^{2})\left[\frac{29}{6}-\ln\left(-\frac{p^2}{\lambda^2}\right)\right]+\mbox{finite}\right\}
\end{align}

\section{Explicit algorithm to express a multiloop integral as scalar BDI's}
\label{app: proof}
In this appendix we provide an explicit algorithm to rewrite BDI's with Lorentz indexes in terms of scalar ones. For simplicity, we focus only on log-divergent integrals and we set surface terms to zero. As explained in section \ref{sec:rules}, in the course of applying IREG rules to a general massless n-loop amplitude, one may encounters an integral of the type
\begin{align}
\mathcal{I}=\int_{k} G(k,p_{i},\mu^{2}) \ln^{n-1}\left[-\frac{(k^2-\mu^{2})}{\lambda^2}\right], 
\end{align}
where $p_i$ stand for external momenta, and $G$ may contain free Lorentz indexes. For simplicity, we assume that the external momenta only appear in denominators. Considering that the integral above is log-divergent, after applying eq. \ref{ident} as many times as the number of external momenta, one obtains
\begin{align}
\mathcal{I}=\mathcal{I}_{\text{div}} + \mathcal{I}_{\text{fin}},\quad\mbox{where}\quad\mathcal{I}_{\text{div}} = \int_{k} G(k,p_{i}=0,\mu^{2}) \ln^{n-1}\left[-\frac{(k^2-\mu^{2})}{\lambda^2}\right].
\end{align}
To conclude the IREG program, one has to write $\mathcal{I}_{\text{div}}$ in terms of BDI's. Therefore, an explicit form for G is needed which, for the sake of generality, we adopt to be
\begin{align}
\mathcal{I}_{\text{div}} = A\int_{k} \frac{k_{\nu_{1}}\cdots k_{\nu_{m}}}{(k^{2}-\mu^{2})^\frac{m+2}{2}} \ln^{n-1}\left[-\frac{(k^2-\mu^{2})}{\lambda^2}\right] = A \left[I_{log}^{(n)}(\mu^{2})\right]_{\nu_{1}\cdots\nu_{m}},
\end{align}
where $A$ is a constant, and $m$ is even. By setting the surface term below to zero,
\begin{align}
\int_k\frac{\partial}{\partial k_{\nu_{1}}}\frac{k^{\nu_{2}}\cdots k^{\nu_{m}}}{(k^{2}-\mu^{2})^\frac{m+2}{2}}\ln^{n-1}\left[-\frac{(k^{2}-\mu^{2})}{\lambda^{2}}\right] = 0,   
\end{align}
one obtains the relation
\begin{align}
\left[I_{log}^{(n)}(\mu^{2})\right]_{\nu_{1}\cdots\nu_{m}} &= \frac{1}{m+2}\left\{g_{\nu_{1}\nu_{2}}\left[I_{log}^{(n)}(\mu^{2})\right]_{\nu_{3}\cdots\nu_{m}} + \cdots g_{\nu_{1}\nu_{m}}\left[I_{log}^{(n)}(\mu^{2})\right]_{\nu_{2}\cdots\nu_{m-1}}\right\}\nonumber\\
&+ \frac{2(l-1)}{m+2}\left[I_{log}^{(n-1)}(\mu^{2})\right]_{\nu_{1}\cdots\nu_{m}}.
\label{eq:bdi}
\end{align}
Notice that by using eq. \ref{eq:bdi}, one reduces a BDI of n-loop order with $m$ free Lorentz indexes to a BDI with two less free Lorentz indexes. Also, one relates a BDI of n-loop order to a BDI one order below. Therefore, by successive applications of eq. \ref{eq:bdi}, one can write $\mathcal{I}_{\text{div}}$ in terms of scalar BDI's only. The end result is not particularly enlightening, thus we do not present it here. However, the coefficient of the $I_{log}^{(n)}(\mu^{2})$ term is easily obtained, which gives
\begin{align}
\mathcal{I}_{\text{div}} = \frac{A}{(m+2)!!}\; g_{\left\{\nu_{1}\nu_{2}\cdots\nu_{m-1}\nu_{m}\right\}}\;I_{log}^{(n)}(\mu^{2}) + \cdots 
\end{align}
where $g_{\left\{\nu_{1}\nu_{2}\cdots\nu_{m-1}\nu_{m}\right\}}$ represents the symmetric combination of all indexes. As can be seen, one obtains the same coefficient for the higher order BDI, regardless of the actual value of n. A similar reasoning can be applied for integrals with higher surface degree of divergence, allowing us to write in general
\begin{align}
\int_{k} G(k,p_{i},\mu^{2}) \ln^{n-1}\left[-\frac{(k^2-\mu^{2})}{\lambda^2}\right] = \mathcal{A}\;I_{log}^{(n)}(\mu^{2})+\cdots
\end{align}
where $\mathcal{A}$ is independent of the actual value of $n$.

\acknowledgments

~M.S. acknowledges a research grant from CNPq (Conselho Nacional de Desenvolvimento Cient\'\i fico e Tecnol\'ogico - 303482/2017-6). We acknowledge support from Fundação para a Ciência e Tecnologia (FCT) through the projects  UID/FIS/04564/2020 and  CERN/FIS-COM/0035/2019. This publication is based upon work from COST Action CA16201 PARTICLEFACE, supported by COST (European Cooperation in Science and Technology, www.cost.eu). This study was financed in part by the Coordenação de Aperfeiçoamento de Pessoal de Nível Superior – Brasil (CAPES) – Finance Code 001.

\end{document}